\documentclass[aps,preprint,floatfix,superscriptaddress,nofootinbib,showpacs]{revtex4}

\usepackage{graphicx}
\usepackage{subfigure}
\usepackage{color}
\usepackage{amsmath}
\usepackage{amssymb}
\usepackage{appendix}
\newcommand{\tbbc}{t \to b \, \bar b \, c}

\newcommand{\bbar}{\overline{b}}
\newcommand{\Lnew}{\Lambda_{\mbox{\scriptsize{new}}}}
\newcommand{\Leff}{{\cal L}_{\mbox{\scriptsize{eff}}}}
\newcommand{\Ahat}{\hat A_i^{\sigma}}
\newcommand{\XIAB}{X^I_{AB}}
\newcommand{\XVLL}{X^V_{LL}}

\newcommand{\abp}{\hat A_b^+}
\newcommand{\abm}{\hat A_b^-}
\newcommand{\abbp}{\hat A_{\bar b}^+}
\newcommand{\abbm}{\hat A_{\bar b}^-}
\newcommand{\acp}{\hat A_c^+}
\newcommand{\acm}{\hat A_c^-}

\newcommand{\Vtb}{V_{tb}}
\newcommand{\Vcb}{V_{cb}}
\newcommand{\MW}{M_W}
\newcommand{\GW}{\Gamma_W}
\newcommand{\mt}{m_t}

\newcommand{\GF}{G_F}

\newcommand{\madg}{\textsc{MadGraph5}}
\newcommand{\feynr}{\textsc{FeynRules}}

\newcommand{\reduce}{\setlength{\baselineskip}{0.75\baselineskip}}

\usepackage{array}
\newcolumntype{C}[1]{>{\centering\let\newline\\\arraybackslash\hspace{0pt}}m{#1}}

\begin{document}

\bibliographystyle{apsrev}

\title{\boldmath Single-top production and rare top interactions}  

\def\hri{\affiliation{\it Harish-Chandra Research Institute, Chhatnag Road, Jhunsi, 
    Allahabad - 211019, India}}
\def\tayloru{\affiliation{\it Physics and Engineering Department,
    Taylor University, \\ 236 West Reade Ave., Upland, IN 46989, USA}}
\def\laplata{\affiliation{\it IFLP, CONICET -- Dpto. de F\'{\i}sica,
    Universidad Nacional de La Plata, C.C. 67, 1900 La Plata,
    Argentina}}

\hri    
\tayloru
\laplata

\author{Pratishruti Saha}
\email{pratishrutisaha@hri.res.in}
\hri

\author{Ken Kiers}
\email{knkiers@taylor.edu}
\tayloru

\author{Alejandro Szynkman}
\email{szynkman@fisica.unlp.edu.ar}
\laplata

\date{\today}

\begin{abstract}
The study of the top quark's properties is an important part of the
LHC programme. In earlier work, we have studied the rare decay $t \to
b \bar b c$, using effective operators to capture the effects of
physics beyond the Standard Model. However top decay is primarily
sensitive to new physics in the sub-TeV energy regime. If this new
physics resides at a higher energy scale, then one needs to turn to
single-top production. In this paper, we use the $s$-channel and $t$-channel 
single-top production measurements to constrain the new physics parameter space
associated with such contact interactions. We also
study the net top polarization as a means to distinguish between
contributions from operators involving different fermion chiralities
and Lorentz structures.
\end{abstract}


\maketitle

%%%%%%%%%%%%%%%%%%%%%%%%%%%%%%%%%%%%%%%%%%%%%%%%%%%%%%%%%%%%%%%%%%%%%%
\section{Introduction}
\label{sec:intro}
%%%%%%%%%%%%%%%%%%%%%%%%%%%%%%%%%%%%%%%%%%%%%%%%%%%%%%%%%%%%%%%%%%%%%%

The top quark has long been believed to be colluding with new
physics (NP). However, intense scrutiny of the top quark's properties at
the Tevatron and at the LHC has so far not revealed any conclusive
departures from the Standard Model (SM). For a few years, the Tevatron
experiments reported a large forward-backward asymmetry in top
pair-production. With the accumulation of more statistics and improved
calculation of the SM predictions, however, this anomaly
disappeared~\cite{Aaltonen:2017efp}. Nevertheless, the top quark
remains a likely suspect -- its mass differs from those of other SM
fermions by orders of magnitude, so much so that it threatens to push
the electroweak vacuum beyond the edge of stability. Moreover,
several anomalies in the $B$-sector~\cite{RK,P5,phimumu,RD} continue 
to fuel speculations that third generation quarks may be the
much-sought window to physics beyond the SM.

In earlier work~\cite{tbbc_sofar}, we proposed a study of rare decay
modes of the top quark. Since all the top quark measurements to
date have been made in channels involving the dominant decay modes of
the top, they would, naturally, have missed signs of new physics that only 
manifests in the rare decay modes. We examined the sensitivity of the LHC 
to the rare decay $\tbbc$ and found that with 3000 fb$^{-1}$ of data, 
the LHC would be able to set statistically significant limits on such decays. 
However, it is evident that top decay would be most sensitive to new physics effects
arising at the energy scale of a few hundred GeV at most. If the
new physics contributions originate at higher energy scales, the
impact on top decay would be too small to be discernible. In order to
probe such interactions further, one must turn to single-top 
production.

In Ref.~\cite{tbbc_sofar}, NP contributions to the 
rare decay $\tbbc$ were parametrized in terms of various four-Fermi 
operators. In this paper, we examine the impact of that same set of 
operators on single-top production. While the top quark decay in the 
mode $\tbbc$ has not been observed, single-top production has been 
measured and these measurements can, in principle, be used to constrain
the strength of these interactions.
\textit{A priori}, it may seem that the contribution of such operators
to single-top production would be diminished by the parton densities
of the heavy quarks in the initial state. While this is true in
general, the situation is salvaged somewhat by the fact that the
competing SM mode is driven by electroweak interactions and
not by strong interactions. A detailed numerical study shows that it
is possible to set meaningful limits on the parameters of the
interactions using existing LHC data. We also present a futuristic
scenario in which very stringent limits may be obtained. 
This possibility, however,
is contingent upon the development of reliable techniques to determine
the charge of an outgoing $b$-quark on an event-by-event basis. We
further examine the possibility of distinguishing between the
contributions of new physics operators with different Lorentz structures
and fermion chiralities using the polarization of the top quark.

Single top production has received significant attention as a direct probe 
of physics beyond the SM~\cite{stp_old,stp_new}.
Since the present analysis focusses on the 
effects of four-quark operators, it is worth mentioning that 
such operators have been studied quite extensively, particularly 
in the context of flavor-changing neutral currents involving the top 
quark~\cite{four-fermion}.

The remainder of this paper is organised as follows. In Section II, 
we discuss the theoretical generalities related to single top production 
at the LHC and introduce the effective operators that we use to parametrize 
new physics contributions, though the detailed analytic expressions are 
consigned to the Appendix. In Section III, we discuss our numerical analyses 
and results for $s$-channel as well as $t$-channel single-top production 
at the LHC at center-of-mass energies of 8 TeV and 13 TeV. We conclude in Section IV.

%%%%%%%%%%%%%%%%%%%%%%%%%%%%%%%%%%%%%%%%%%%%%%%%%%%%%%%%%%%%%%%%%%%%%%
\section{Single top production}
\label{sec:theory}
%%%%%%%%%%%%%%%%%%%%%%%%%%%%%%%%%%%%%%%%%%%%%%%%%%%%%%%%%%%%%%%%%%%%%%

\begin{figure}[!htbp]
\subfigure[]{\includegraphics[scale=0.35]{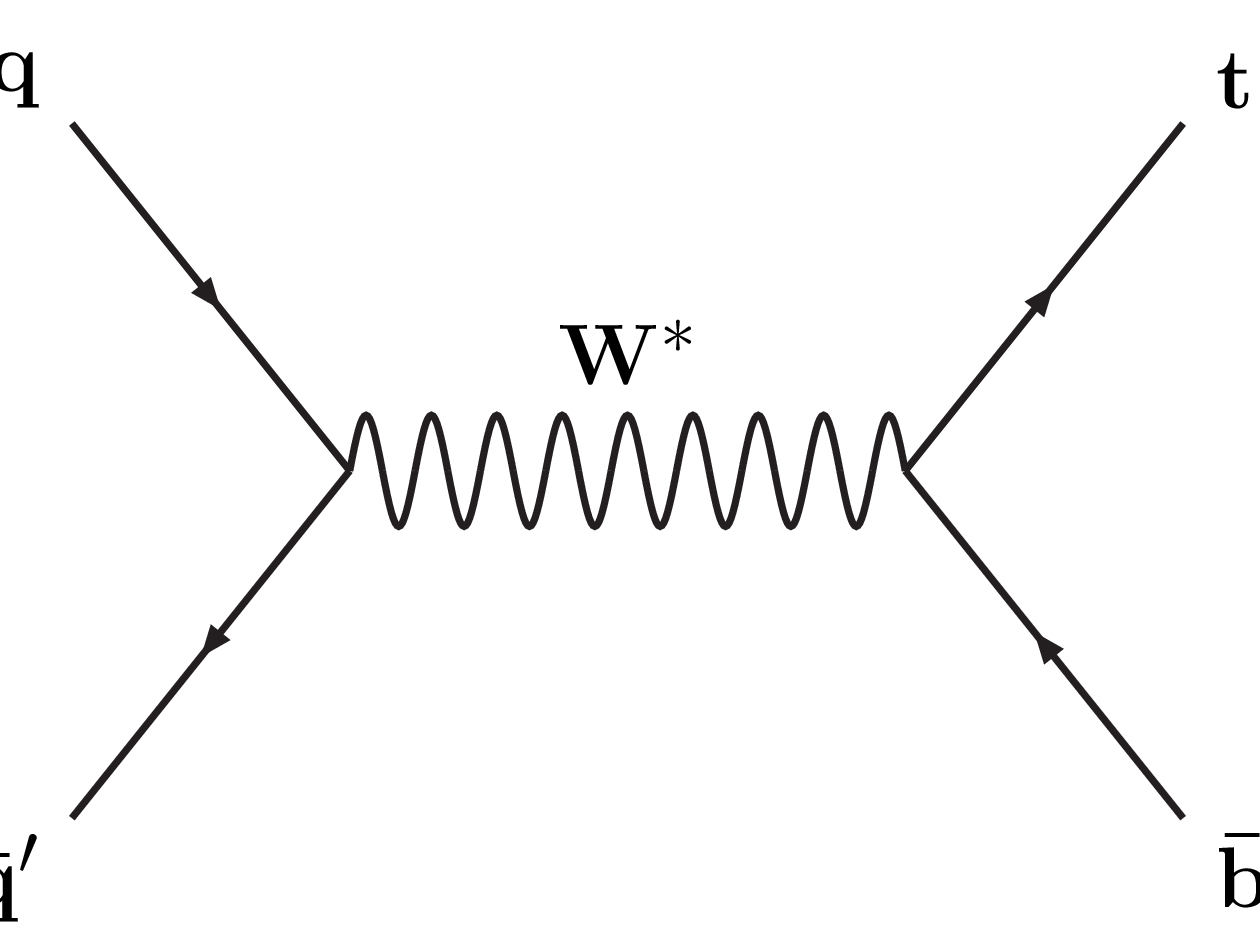}}
\hspace*{70pt}
\subfigure[]{\includegraphics[scale=0.35]{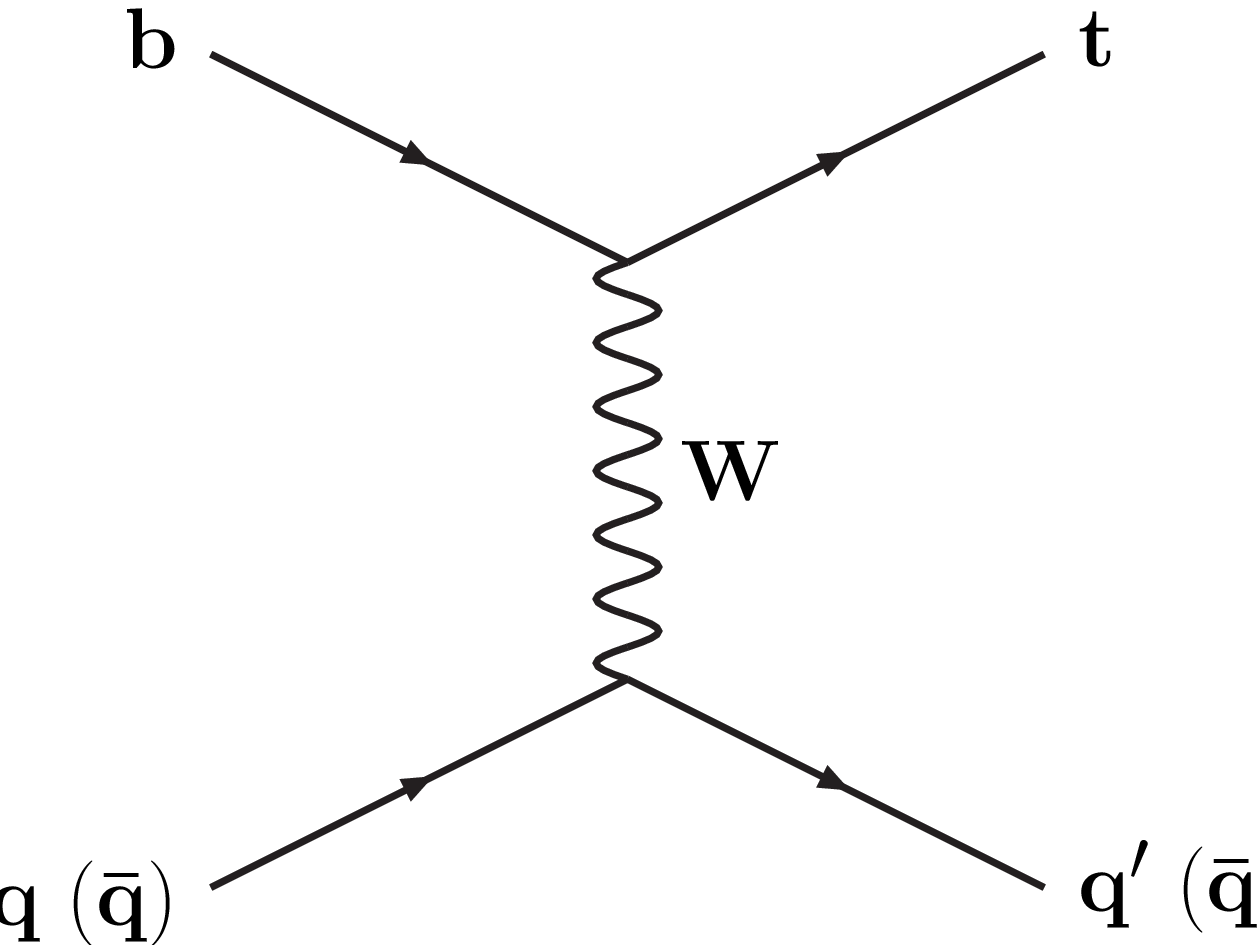}} \\
\subfigure[]{\includegraphics[scale=0.35]{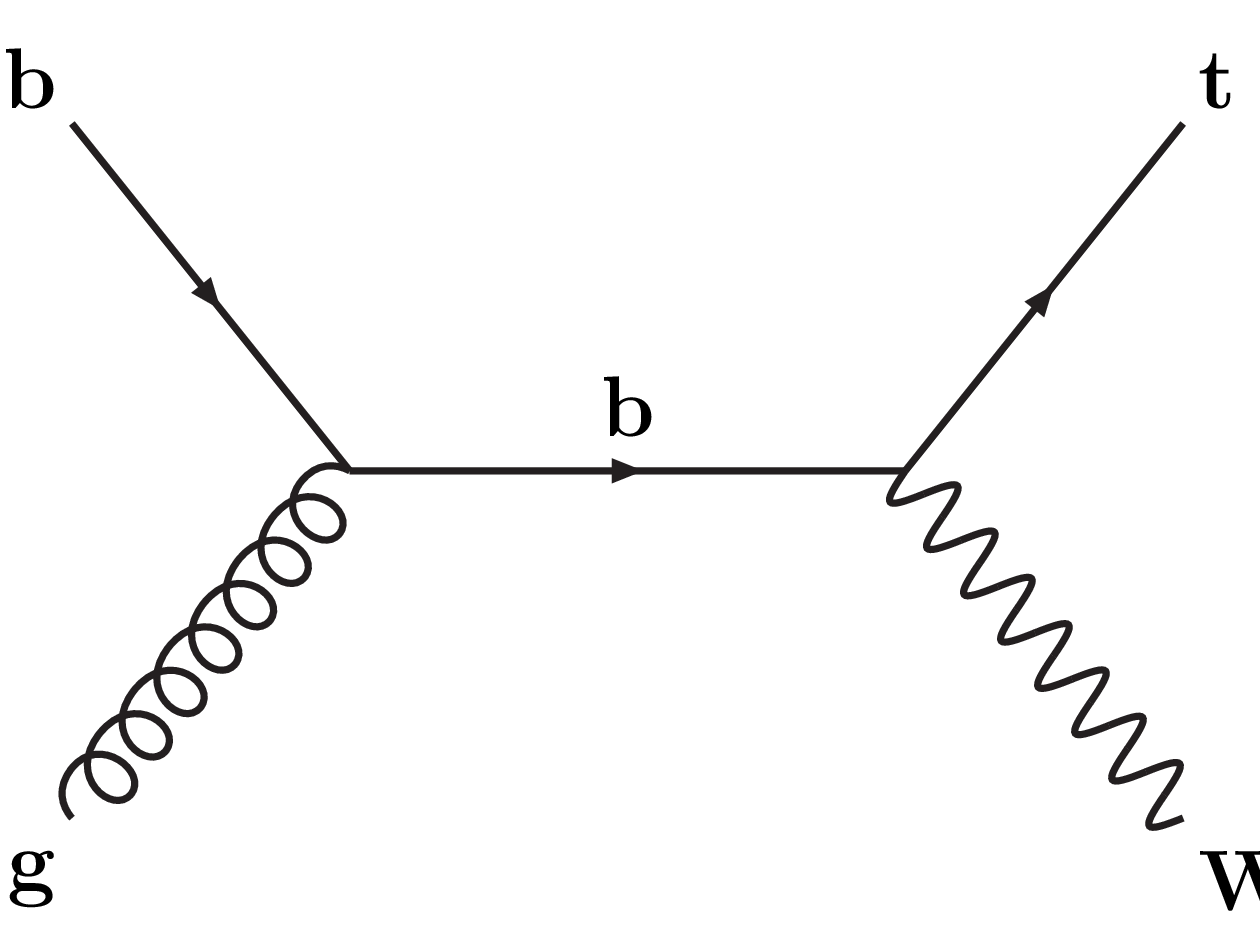}}
\hspace*{70pt}
\subfigure[]{\includegraphics[scale=0.35]{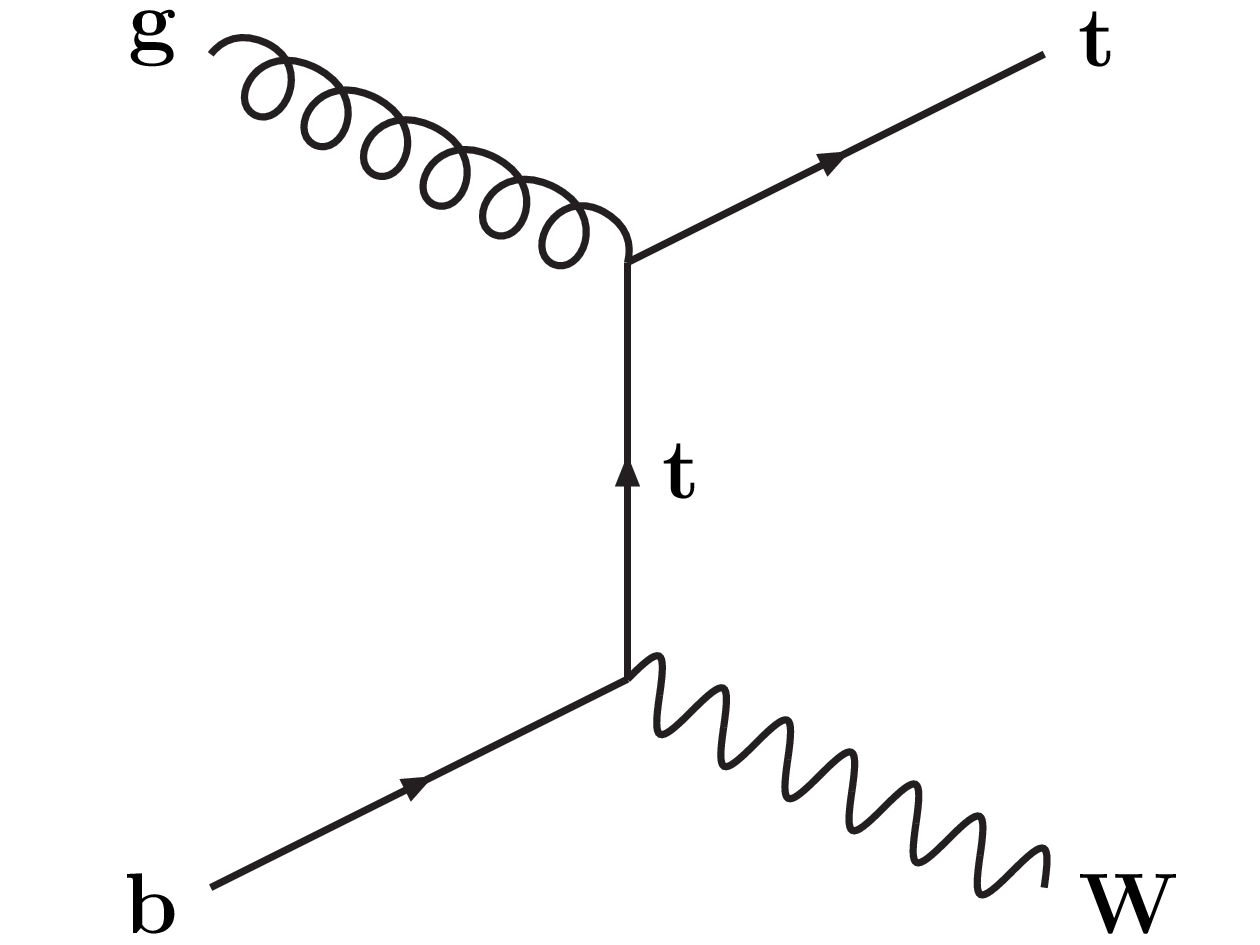}}
\caption{(a) $s$-channel; (b) $t$-channel; (c) \& (d) $Wt$-channel single-top production
  modes.}
\label{fig:production_modes}
\end{figure}

\begin{figure}[!htbp]
\includegraphics[scale=0.45,angle=-90]{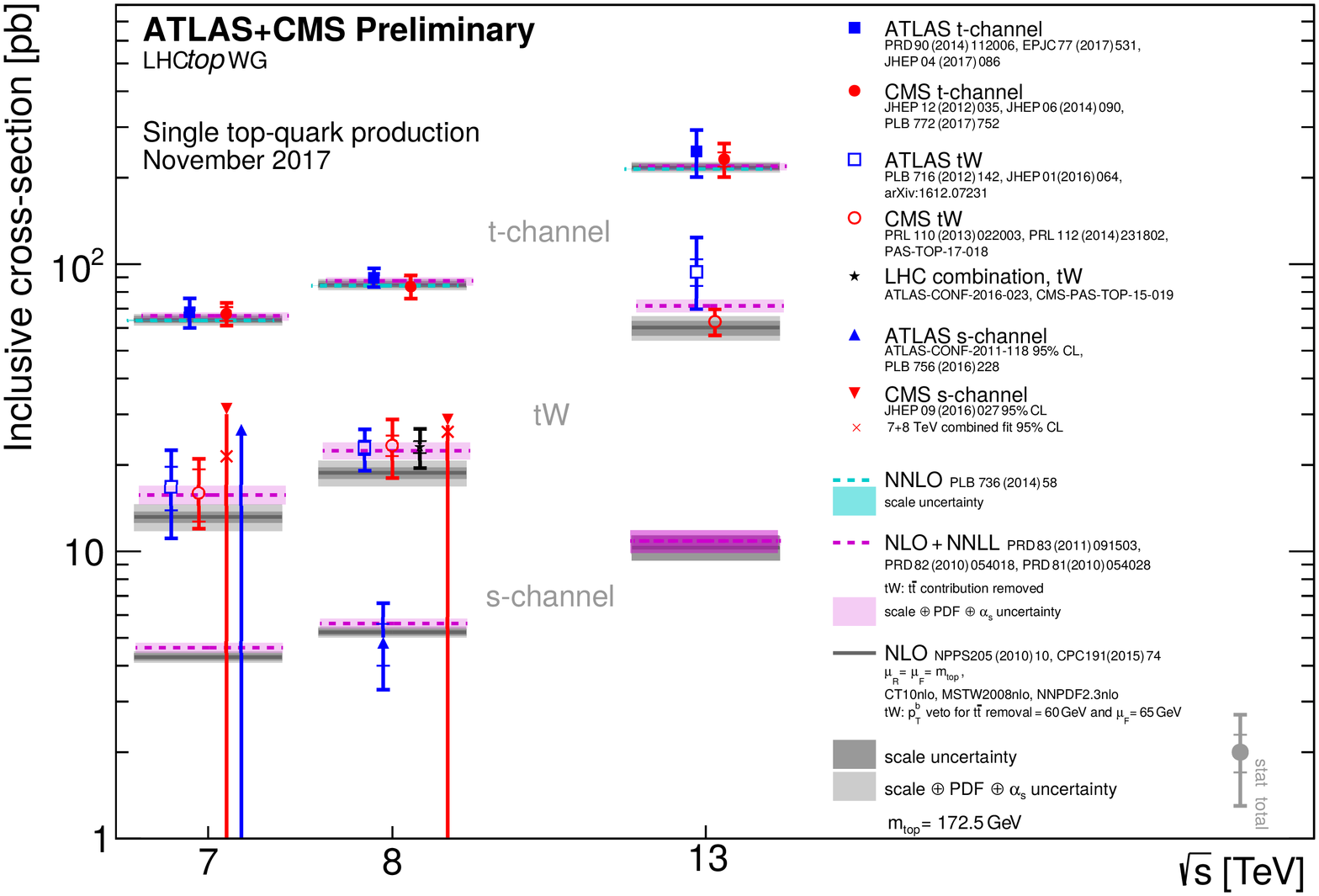}
\caption{Summary of single-top production rates measured at the LHC~\cite{LHC_top_WG}.}
\label{fig:stp_LHC_summary}
\end{figure}

At a hadron collider, the dominant production mode for top quarks is
$pp \, (p \bar p) \to t \bar t$.  Single top production is
sub-dominant. Nonetheless, it is important as it provides a cleaner
way of measuring the electroweak couplings of the top quark. Within
the framework of the SM, single-top production at hadron
colliders is classified into 3 production channels as shown in
Fig.~\ref{fig:production_modes}, namely, $s$-channel, $t$-channel and
$Wt$-channel. At the LHC, $t$-channel production is the dominant mode,
followed by $Wt$ associated production. Cross sections for all three
channels have been measured at the LHC during the 7-TeV and 8-TeV
runs~\cite{ATLAS_t7,ATLAS_t8,ATLAS_s7,ATLAS_s8,CMS_t7,CMS_t8,CMS_s7n8}. The
13-TeV run is ongoing and results are already available in some
channels~\cite{ATLAS_t13,CMS_t13}. These are summarised in
Fig.~\ref{fig:stp_LHC_summary}.
For $t$-channel single top production some kinematic 
distributions have also been measured~\cite{ATLAS_t8,CMS_dists_8,CMS_dists_13}.
All the measurements are, by-and-large, in agreement with the SM predictions, even
though certain channels are plagued by large experimental
uncertainties.

\vspace*{40pt}

\subsection{Effective Lagrangian}
\label{sec:Leff}
\vspace*{-15pt}
In Ref.~\cite{tbbc_sofar}, the contributions from physics beyond the
SM to the rare decay $\tbbc$ were parametrized in terms
of the 6-dimensional operators given by $\Leff$,
\[
{\cal L}_{\mbox{\scriptsize eff}} \quad = \quad  
{\cal L}_{\mbox{\scriptsize eff}}^V \;\;+\;\;
{\cal L}_{\mbox{\scriptsize eff}}^S \;\;+\;\; 
{\cal L}_{\mbox{\scriptsize eff}}^T ,\]
where
\begin{eqnarray}
  {\cal L}_{\mbox{\scriptsize eff}}^V & = & 4\sqrt{2}G_F V_{cb}V_{tb}
       \left\{
    X_{LL}^V\,\bbar\gamma_\mu P_L t \,
       \overline{c}\gamma^\mu P_L b
   + X_{LR}^V\,\bbar\gamma_\mu P_L t \,
       \overline{c}\gamma^\mu P_R b
\right.\nonumber\\
& & \hskip2.2truecm \left.
   +~X_{RL}^V\,\bbar\gamma_\mu P_R t \,
       \overline{c}\gamma^\mu P_L b
   + X_{RR}^V\,\bbar\gamma_\mu P_R t \,
       \overline{c}\gamma^\mu P_R b
\right\}+ \mbox{h.c.}, 
\label{eq:eff1}\\
&&\nonumber\\
  {\cal L}_{\mbox{\scriptsize eff}}^S & = & 4\sqrt{2}G_F V_{cb}V_{tb}
\left\{
     X_{LL}^S\,\bbar P_L t \,\overline{c} P_L b
   + X_{LR}^S\,\bbar P_L t \,\overline{c} P_R b
\right. \nonumber\\
& & \hskip2.2truecm \left.
   +~X_{RL}^S\,\bbar P_R t \,\overline{c} P_L b
   + X_{RR}^S\,\bbar P_R t \,\overline{c} P_R b
\right\}+\mbox{h.c.,} 
\label{eq:eff2}\\
&&\nonumber\\
  {\cal L}_{\mbox{\scriptsize eff}}^T & = & 4\sqrt{2}G_F V_{cb}V_{tb}
\left\{
     X^T_{LL} \overline{b}\sigma^{\mu\nu}P_L t \,
     \overline{c}\sigma_{\mu\nu}P_L b 
\right. \nonumber\\
& & \hskip2.2truecm \left.     +~X^{T}_{RR}
\bbar\sigma^{\mu\nu}P_R t \,
     \overline{c}\sigma_{\mu\nu} P_R b
\right\}+\mbox{h.c.}
\label{eq:eff3}
\end{eqnarray}

\noindent
Clearly, these operators can also contribute to single-top
production. The $Wt$-channel would remain unaffected by these NP
contributions. However, final states identical to those produced in
$s$-channel and $t$-channel processes can occur via the contact interactions
listed above. 

\subsection{Contribution to single-top production from $\Leff$}
\label{sec:Leff_contribution}

The operators listed in Eqs.~(\ref{eq:eff1}), (\ref{eq:eff2}) and
(\ref{eq:eff3}) can give rise to three possible amplitudes for single
top production : $\bar b c \to t \bar b$, $b c \to t b$, $b \bar b
\to t \bar c$.  In the SM, the first one is an $s$-channel
process, whereas the second and third are $t$-channel processes. In
addition, the three final states get contributions from light-quark
initial states in the SM. Some of the key features are as follows:

\begin{itemize}
\item[(i)] For single-top production due to such operators, the
  initial states would necessarily consist of bottom and charm quarks. 
  The low densities of these inside the proton tend
  to suppress the cross section as compared to the SM production
  rates. This effect is more pronounced in the $s$-channel than in the
  $t$-channel as the SM rate for the latter is CKM supressed unless
  there is a $b$ quark in the initial state (see Fig.~\ref{fig:production_modes}).
           
\item[(ii)] The suppression caused by the low initial-state parton
  densities is compensated to some extent by the growth in the
  cross section with increase in the parton c.m. energy ($\sqrt{\hat
    s}$). This, of course, is a generic feature of contact
  interactions.
             
\item[(iii)] When new physics is parametrized in terms of effective
  operators, the implicit assumption is that the operators arise due
  to physics at a very high energy scale ($\Lnew$) that is beyond the
  direct reach of current experiments. When the operators listed in
  Eqs.~(\ref{eq:eff1})-(\ref{eq:eff3}) are applied to top decay,
  $\Lnew \sim {\cal{O}}$(TeV) is permissible, given that the energy
  scale of the interaction is $m_t \sim $173 GeV. However, in the
  context of single-top production in $pp$ collisions at LHC energies,
  one must necessarily consider significantly larger values of $\Lnew
  \sim {\cal{O}}$(10 TeV) or higher. In the parametrization used in
  this work and in our previous work on this topic~\cite{tbbc_sofar},
  $\Lnew \sim (G_F \, X^{I}_{AB})^{-1/2}$.  Consequently, $\XIAB \sim
  {\cal O}(1)$ correspond to $\Lnew$ in the sub-TeV regime.  In
  principle, it should be possible to rule out such new physics
  scenarios as they could lead to resonances in the $s$-channel
  production mode, causing a spike in the cross section.  However,
  owing to the relatively large uncertainty in the measured
  $s$-channel cross section, no robust conclusions can be drawn at
  this stage. In the $t$-channel, such contributions would manifest, 
  for example, in the form of a harder transverse momentum distribution.
  However, no such deviations have been found up to $p_T \sim$ 
  300 GeV~\cite{ATLAS_t8,CMS_dists_8,CMS_dists_13}. 
\end{itemize}

In our previous work related to $\tbbc$, the 10 couplings $\XIAB$ from
the NP effective Lagrangian $\Leff$ were found to appear together in
six characteristic combinations.  We denoted these six combinations by
$\Ahat$, defining

\vspace*{-10pt}

\begin{align}
  \hat{A}_{\bbar}^+ \;\; &= \;\; 4 \left|X^{V}_{LL}\right|^2
  -8 \,\mbox{Re}\left(X^T_{LL}X^{S*}_{LL}\right)+32 \left|X^T_{LL}\right|^2 ~,
  \mspace{100mu} \nonumber\\
  \hat{A}_{\bbar}^- \;\; &= \;\;
    4\left|X^{V}_{RR}\right|^2
    -8 \,\mbox{Re}\left(X^T_{RR}X^{S*}_{RR}\right)+32 \left|X^T_{RR}\right|^2 ~, \nonumber \\
  \hat{A}_{b}^+ \;\; &= \;\; 
    \left|X^{S}_{LL}\right|^2+\left|X^{S}_{LR}\right|^2
    -16\left|X^{T}_{LL}\right|^2 ~, \nonumber \displaybreak \\
  \hat{A}_{b}^- \;\; &= \;\; 
    \left|X^{S}_{RR}\right|^2+\left|X^{S}_{RL}\right|^2
    -16\left|X^{T}_{RR}\right|^2 ~,\nonumber \\
  \hat{A}_{c}^+ \;\; &= \;\; 
    4\left|X^{V}_{LR}\right|^2
    +8 \,\mbox{Re}\left(X^T_{LL}X^{S*}_{LL}\right)+32\left|X^T_{LL}\right|^2 ~, \nonumber\\
  \hat{A}_{c}^- \;\; &= \;\; 
    4\left|X^{V}_{RL}\right|^2
    +8 \,\mbox{Re}\left(X^T_{RR}X^{S*}_{RR}\right)+32\left|X^T_{RR}\right|^2 ~.
\label{eq:Ahatdefs}
\end{align}
In addition to depending on the above six quantities, various
observables were also found to depend on the real and imaginary parts
of $X^V_{LL}$ and on the combination
$\mbox{Im}\left(X^T_{LL}X^{S*}_{LL}+X^T_{RR}X^{S*}_{RR}\right)$.\footnote{The
  dependence on $X^V_{LL}$ comes from SM-NP interference terms.  Terms
  proportional to
  $\mbox{Im}\left(X^T_{LL}X^{S*}_{LL}+X^T_{RR}X^{S*}_{RR}\right)$ are
  associated with triple-product correlations~\cite{tbbc_sofar}.}  
  
In the case of single-top production, once again, NP contributions to the various cross
sections of interest can be expressed in terms of these same
combinations of NP parameters.  Explicit expressions for the matrix
elements squared in the different cases may be found in the
Appendix. If we restrict our attention to the case in which we sum
over the top quark's spin, we find that the 20-dimensional parameter
space spanned by the 10 complex-valued parameters $\XIAB$ is reduced
to a 5-dimensional parameter space composed of $(\abbp + \abbm)$,
$(\abp + \abm)$, $(\acp + \acm)$, Re($X^V_{LL}$) and
Im($X^V_{LL}$).\footnote{As we have noted in past work, the situation
  is somewhat complicated by the fact that the real and imaginary
  parts of $X^V_{LL}$ also appear in the parameter $\abbp$.}  If one wishes
to consider top quark polarization effects, there is an additional
dependence on $(\abbp - \abbm)$, $(\abp - \abm)$, $(\acp - \acm)$, and
$\mbox{Im}\left(X^T_{LL}X^{S*}_{LL}+X^T_{RR}X^{S*}_{RR}\right)$
(please refer to the Appendix for details).  In the present work we
shall assume that $X^V_{LL}$ and
$X^T_{LL}X^{S*}_{LL}+X^T_{RR}X^{S*}_{RR}$ are both real, thereby
reducing the size of the parameter space somewhat.\footnote{Note that
  $\mbox{Im}(X^V_{LL})$ plays an important role in partial rate
  asymmetries~\cite{tbbc_sofar}; we neglect such
  effects here.}  In our numerical work below, we will mostly consider
the case in which the polarization of the top quark is ignored
(Secs.~\ref{sec:t_channel}, \ref{sec:s_channel}, \ref{sec:diff_dist} and \ref{sec:futuristic}). 
A polarization-dependent asymmetry is considered in Sec.~\ref{sec:polarization}.
Furthermore, throughout the remainder of this paper, 
we will consider operators with a single Lorentz structure at a time. We will 
also make certain assumptions about the relative magnitudes of the couplings 
associated with operators having the same Lorentz structure but different chirality 
structure.

We close this section with the following comment.
The careful reader might have noticed that the effective operators 
listed in Eqs.~(\ref{eq:eff1}), (\ref{eq:eff2}) and (\ref{eq:eff3})
are not SM gauge invariant. 
Within a gauge-invariant framework, the presence of new operators, 
potentially constrained by other observables, could limit the size of 
the Wilson coefficients corresponding to the operators in our basis. 
We have examined the gauge-invariant operators involving the second 
and third quark generations and have found that the observables that 
they may directly affect are bottom and top pair production. 
As is the case for single top production within our framework, 
the new operators lead to effects suppressed by the PDFs of the 
initial heavy quarks. However, for bottom and top pair production 
the contributions from the new operators must also compete with 
the dominant QCD contributions.
Therefore we expect these processes to be far less constraining
than those considered in the present work.

%%%%%%%%%%%%%%%%%%%%%%%%%%%%%%%%%%%%%%%%%%%%%%%%%%%%%%%%%%%%%%%%%%%%%%
\section{Numerical Study}
\label{sec:numerical}
%%%%%%%%%%%%%%%%%%%%%%%%%%%%%%%%%%%%%%%%%%%%%%%%%%%%%%%%%%%%%%%%%%%%%%

In order to extract limits on the $\Ahat$'s (and consequently on
$\XIAB$), we start by implementing $\Leff$ alongside the Standard
Model in {\madg}~\cite{MG5} using {\feynr}~\cite{FR}. This puts us in a position to
calculate the tree-level cross section for single-top production in
$pp$ collisions, for which we use CTEQ6L parton distributions
functions (PDFs)~\cite{CTEQ}, setting both the renormalization and factorization
scales to be $\mt$ = 173 GeV. In order to approximate higher-order QCD
corrections, we estimate K-factors as
$\sigma^{SM}_{NNLO}$(approx.)/$\sigma^{SM}_{LO}$, where
$\sigma^{SM}_{NNLO}$(approx.) is obtained from the references
listed in Table~\ref{tab:K-Factor} and $\sigma^{SM}_{LO}$ is
calculated using {\madg} in conjunction with MSTW2008LO
PDFs~\cite{MSTW}.\footnote{MSTW2008LO PDFs are used only for the purpose of
  determining the K-factor, to be consistent with the calculations for
  $\sigma^{SM}_{NNLO}$(approx.). For all subsequent calculations we
  use CTEQ6L PDFs.}  We then compute the tree-level cross sections,
including both SM and NP effects, and multiply the results by the
corresponding K-factors to obtain estimates of the QCD-corrected
values.  In the following, ``$\sigma_{SM}$'' denotes the SM cross
sections obtained in this manner.

\begin{table}[!htbp]
\begin{tabular}{|l|C{3cm}|C{5cm}|C{2cm}|}
\hline
                    & $\sigma_{NNLO}$             & $\sigma_{LO}$                      & $K$  \\
\hline
$s$-channel; 8 TeV  & From Ref.~\cite{NLO_sch_813} & From {\madg} using MSTW2008LO PDFs & 1.74 \\
\hline
$s$-channel; 13 TeV & From Ref.~\cite{NLO_sch_813} & From {\madg} using MSTW2008LO PDFs & 1.73 \\ 
\hline
$t$-channel; 8 TeV  & From Ref.~\cite{NLO_tch_8}   & From {\madg} using MSTW2008LO PDFs & 1.06 \\
\hline
$t$-channel; 13 TeV & From Ref.~\cite{CMS_t13}     & From {\madg} using MSTW2008LO PDFs & 1.02 \\
\hline
\end{tabular}
\caption{\reduce K-factors for $s$- and $t$-channel single-top production
cross sections at $8$ and $13$~TeV.} 
\label{tab:K-Factor}
\end{table}

%%%%%%%%%%%%%%%%%%%%%%%%%%%%%%%%%%%%%%%%%%%%%%%%%%%%%%%%%%%%%%%%%%%%%%
\subsection{$t$-channel single-top production}
\label{sec:t_channel}
\begin{figure}[!htbp]
\includegraphics[scale=0.5]{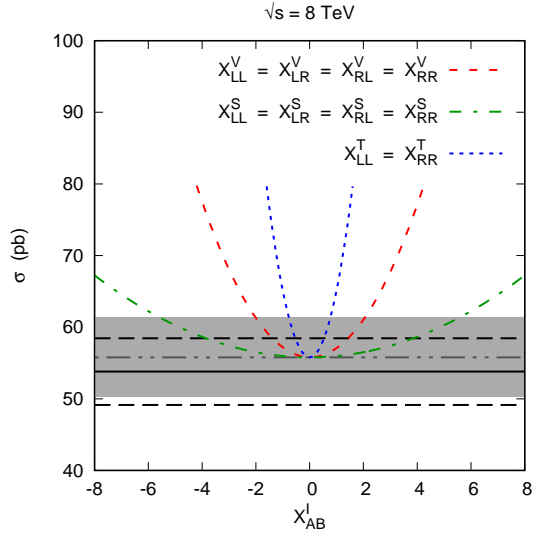}
\vspace*{-10pt}
\caption{\reduce $t$-channel single-top production cross section at
  $\sqrt{s}$ = 8 TeV. The black solid and dashed horizontal lines
  depict the CMS measurement~\cite{CMS_t8} along with the total
  uncertainty. The grey dot-dashed line and the grey shaded region
  reflect $\sigma_{SM} \pm 10\%$, which is calculated as described in
  the text. The red dashed, green dot-dashed and blue dotted curves
  give the cross section in the presence of NP vector, scalar and
  tensor interactions, respectively. As mentioned in the text, only
  one Lorentz structure is considered at a time.}
\label{fig:t-ch_8TeV_all}
\end{figure}

%%%%%%%%%%%%%%%%%%%%%%%%%%%%%%%%%%%%%%%%%%%%%%%%%%%%%%%%%%%%%%%%%%%%%%
\vspace*{-10pt}
At the LHC, $t$-channel processes yield the dominant contribution to
single-top production.  These processes consist of $d_i u_j \to t d_k$
and $d_i \bar d_j \to t \bar u_k$ where $d_{i,j,k} \in \{d,s,b\}$ and 
$u_{j,k} \in \{u,c\}$. Within the SM, the relative magnitudes of the contributions from the different
initial states are governed by the densities of the respective partons
inside the proton and the CKM factors appearing in the amplitude. Once
$\Leff$ is introduced, there are additional contributions to $b c \to
t b$ and $b \bar b \to t \bar c$. Figure~\ref{fig:t-ch_8TeV_all} shows
the cross section. As noted above, the operators in ${\cal L}_V$,
${\cal L}_S$ and ${\cal L}_T$ are considered separately in our
analysis. It can be seen that the tensor operators are constrained
most tightly, followed by vector and scalar operators. This is
expected, given the structure of the $\Ahat$s and the fact that
large numerical factors accompany $X^T_{AB}$ wherever they appear.
It is intriguing to note that, except in the case of $X^T_{AB}$,
couplings of ${\cal O}$(1) are not excluded by the experimental data
at 8 TeV. How do we reconcile this with a) our earlier statement that
$\XIAB \sim {\cal O}$(1) correspond to $\Lnew$ in the sub-TeV range,
and b) the fact that no exotic physics has been discovered at the LHC
so far? We return to this question below, in Sec.\ref{sec:futuristic}.  

\begin{figure}[!htbp]
\includegraphics[scale=0.5]{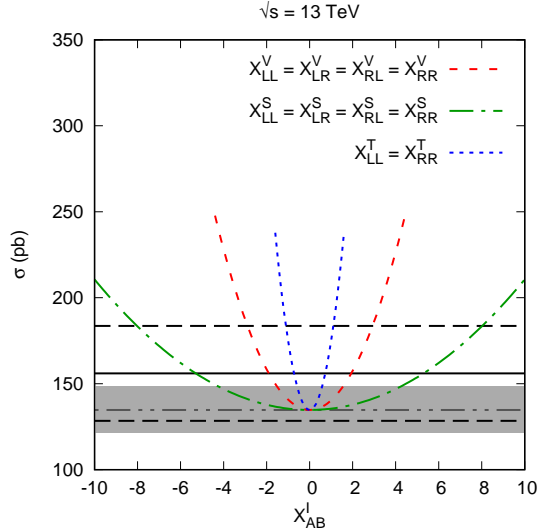}
\vspace*{-10pt}
\caption{\reduce $t$-channel single-top production cross section at
  $\sqrt{s}$ = 13 TeV. The black solid and dashed horizontal lines
  depict the ATLAS measurement~\cite{ATLAS_t13} along with the total
  uncertainty. 
  The other lines and the shaded region have the same meanings
  as in Fig.~\ref{fig:t-ch_8TeV_all}.
%   The grey dot-dashed line and the grey shaded region
%   reflect $\sigma_{SM} \pm 10\%$, which is calculated as described in
%   the text. The red dashed, green dot-dashed and blue dotted curves
%   give the cross section in the presence of NP vector, scalar and
%   tensor interactions, respectively. As noted above, only one Lorentz
%   structure is considered at a time.
}
\label{fig:t-ch_13TeV_all}
\end{figure}

Figure~\ref{fig:t-ch_13TeV_all} shows the
analogous analysis for the 13-TeV data.  As compared to the 8-TeV
data, the 13-TeV data appear to be less constraining. At first glance,
one is tempted to attribute this to low statistics given that the
8-TeV measurement is based on 19.7 fb$^{-1}$ of data while the 13-TeV
result is based on 3.2 fb$^{-1}$. However, despite the relatively low
statistics at 13-TeV, it turns out that the largest component of the
uncertainty is due to systematics. If future analyses can reduce the
systematic uncertainty, then tighter constraints can be expected.
Presently, for a more effective comparison between the sensitivities 
to NP couplings at 8 TeV and 13 TeV, we construct a 10\% band around 
the SM prediction (see Figs.~\ref{fig:t-ch_8TeV_all} and \ref{fig:t-ch_13TeV_all}). 
This gives us an estimate of the improvement in the limits under the assumption 
that, at both 8 TeV and 13 TeV, the central value of the measurement coincides 
with the SM prediction and has a 10\% uncertainty. 

\begin{figure}[!htbp]
\subfigure[]{\includegraphics[scale=0.5]{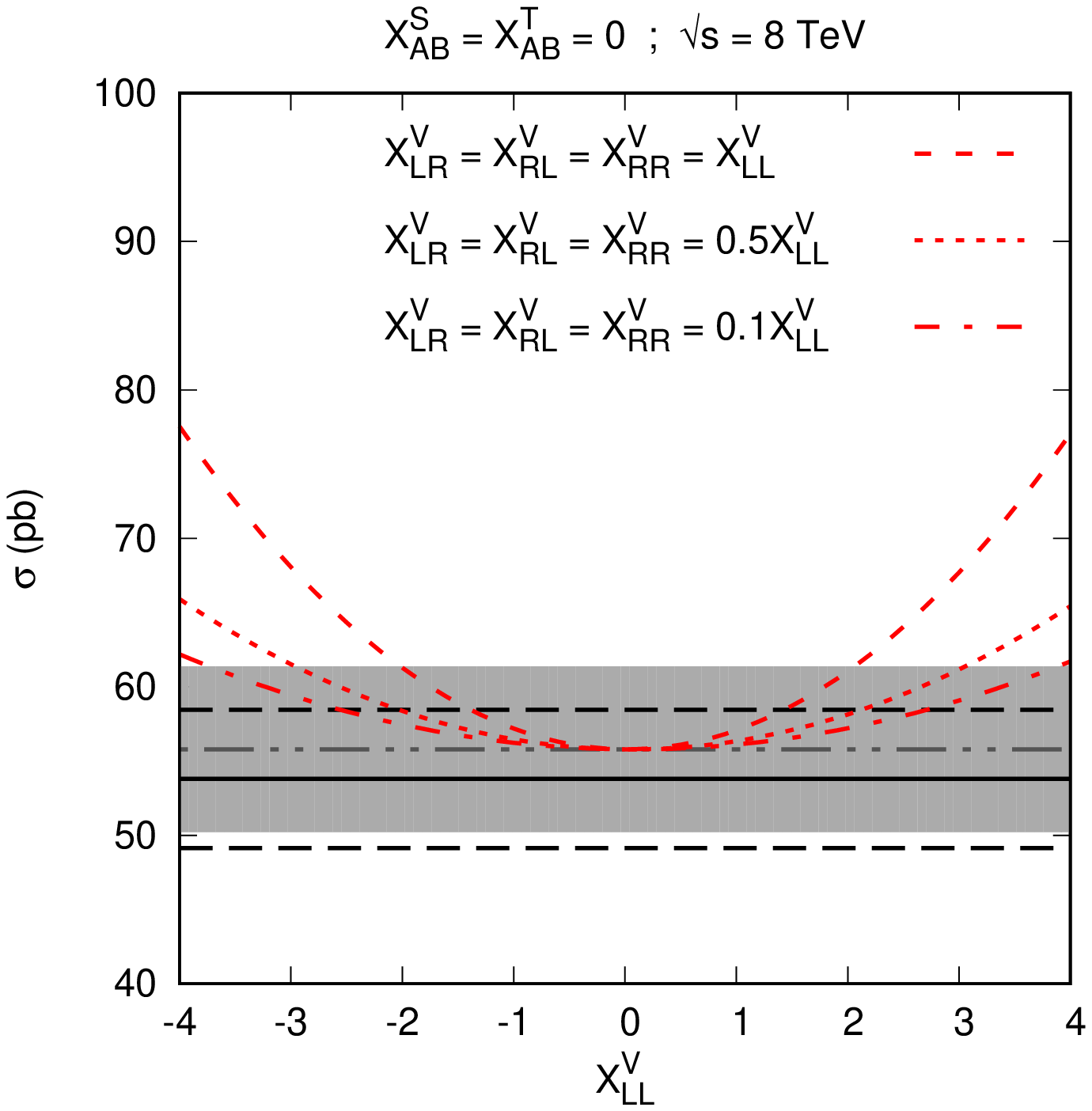}}
\subfigure[]{\includegraphics[scale=0.5]{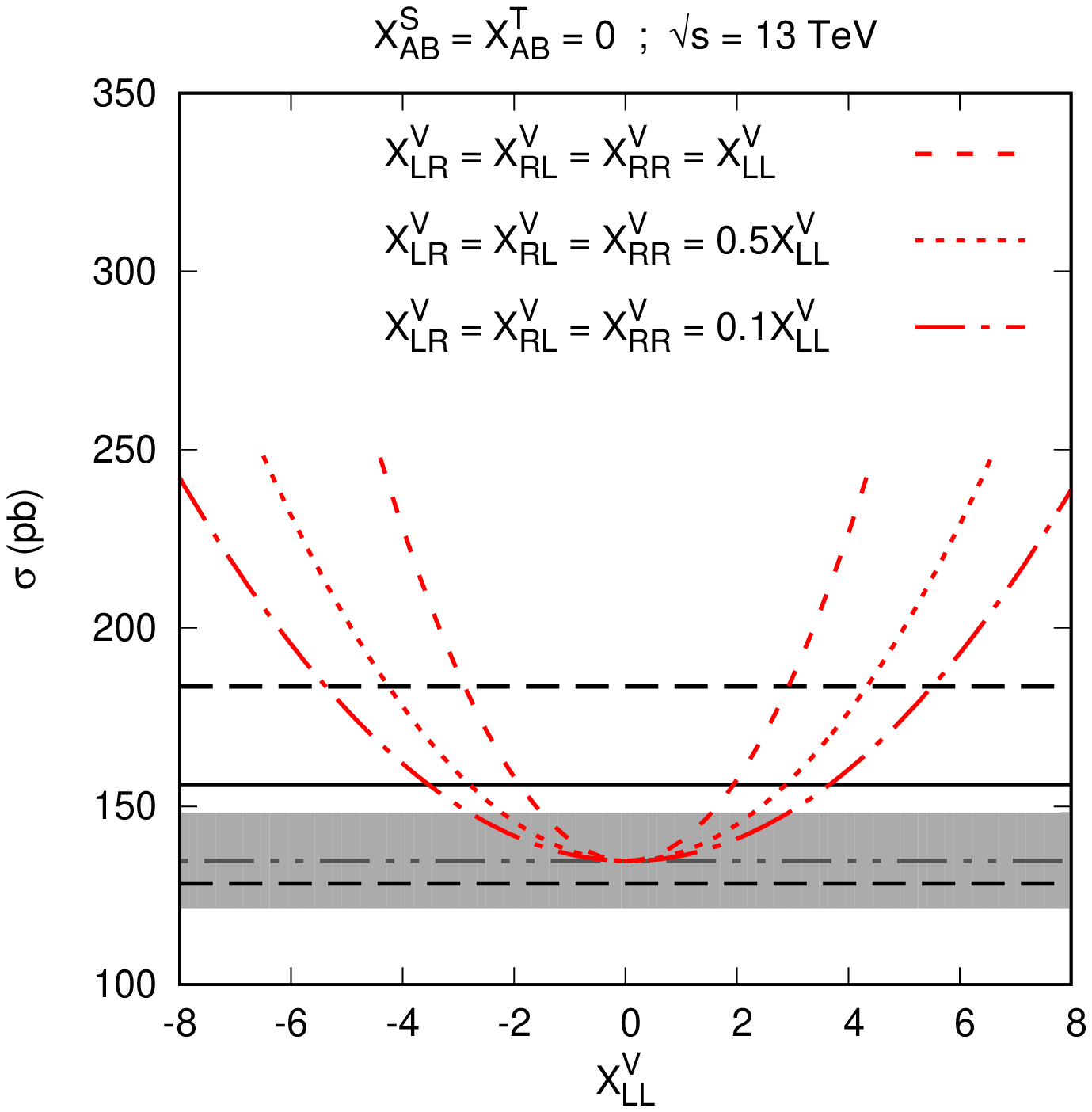}} 
\caption{\reduce $t$-channel single-top production cross section in
  the presence of various combinations of NP vector contributions at
  (a) 8 TeV and (b) 13 TeV.  In the former case, the black solid and
  dashed horizontal lines depict the CMS measurement~\cite{CMS_t8}
  along with the total uncertainty; in the latter case, the black
  solid and dashed horizontal lines depict the ATLAS
  measurement~\cite{ATLAS_t13} along with the total uncertainty.  In
  both cases, the grey band depicts $\sigma_{SM} \pm 10\%$.}
\label{fig:t-ch_V}
\end{figure}

\begin{figure}[!htbp]
\subfigure[]{\includegraphics[scale=0.5]{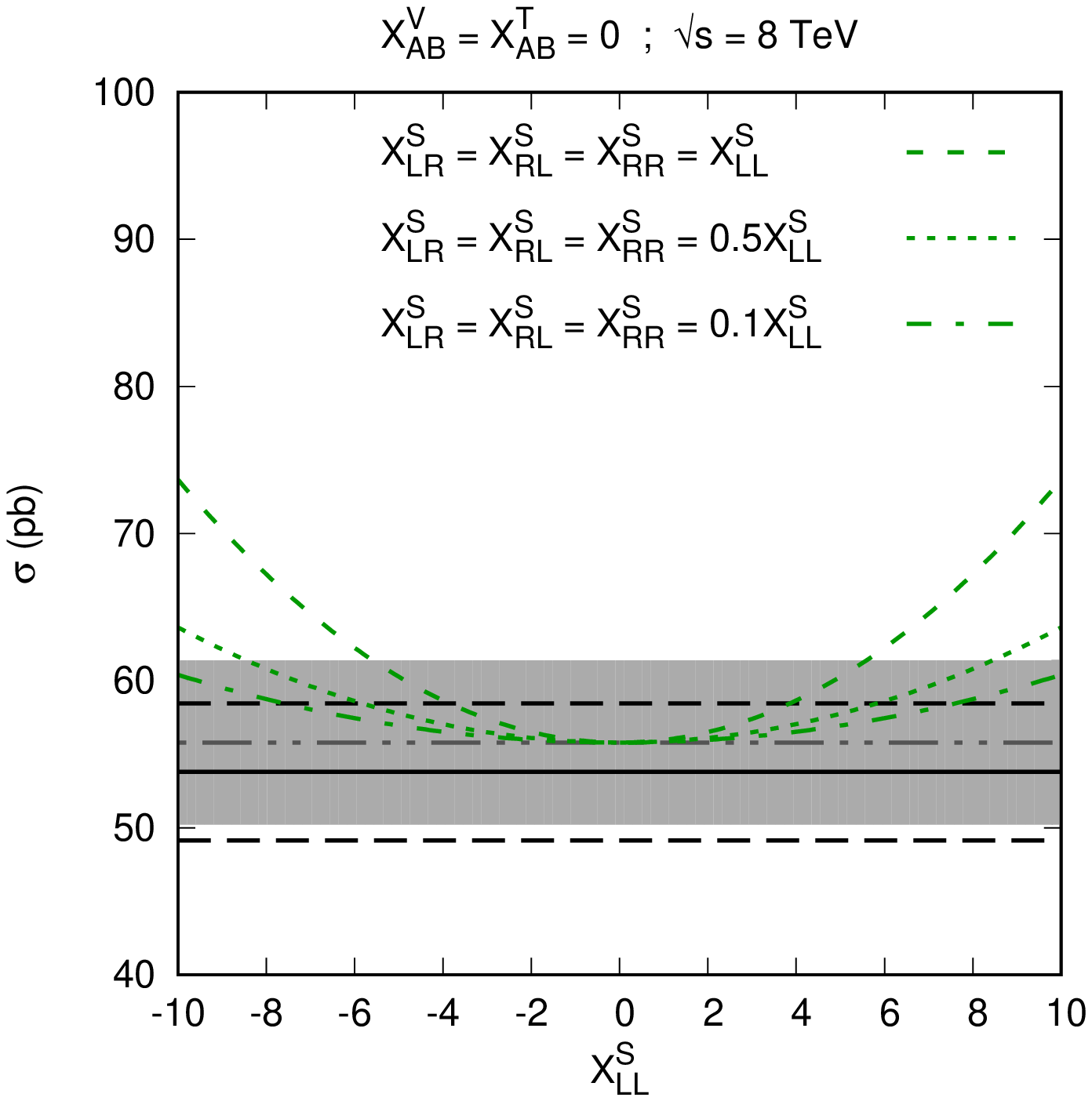}}
\subfigure[]{\includegraphics[scale=0.5]{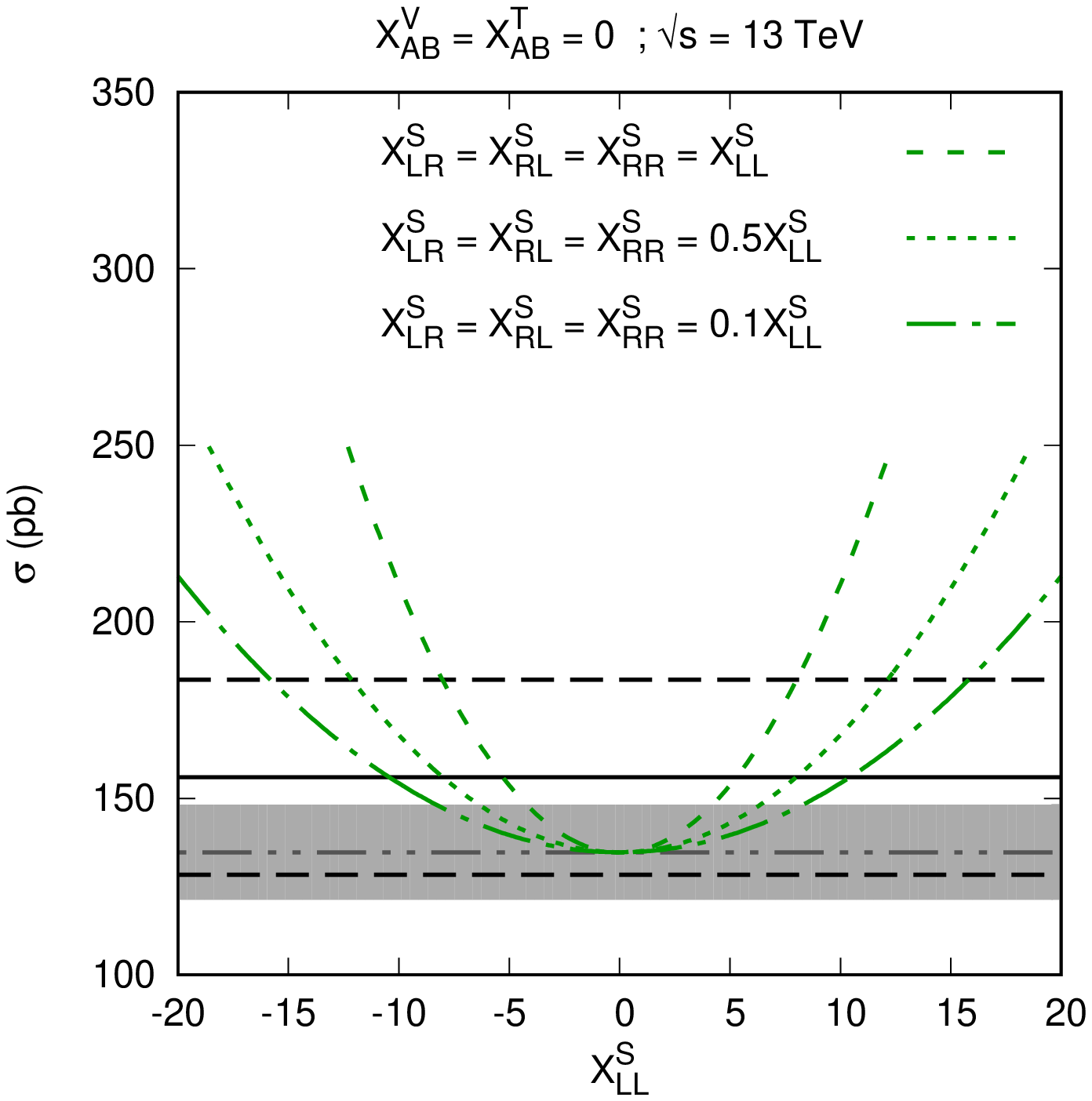}} 
\caption{\reduce $t$-channel single-top production cross section in
  the presence of NP scalar contributions at (a) 8 TeV and (b) 13 TeV.
  The horizontal lines and shaded regions have the same meanings as in
  Fig.~\ref{fig:t-ch_V}.}
\label{fig:t-ch_S}
\end{figure}

\begin{figure}[!htbp]
\subfigure[]{\includegraphics[scale=0.5]{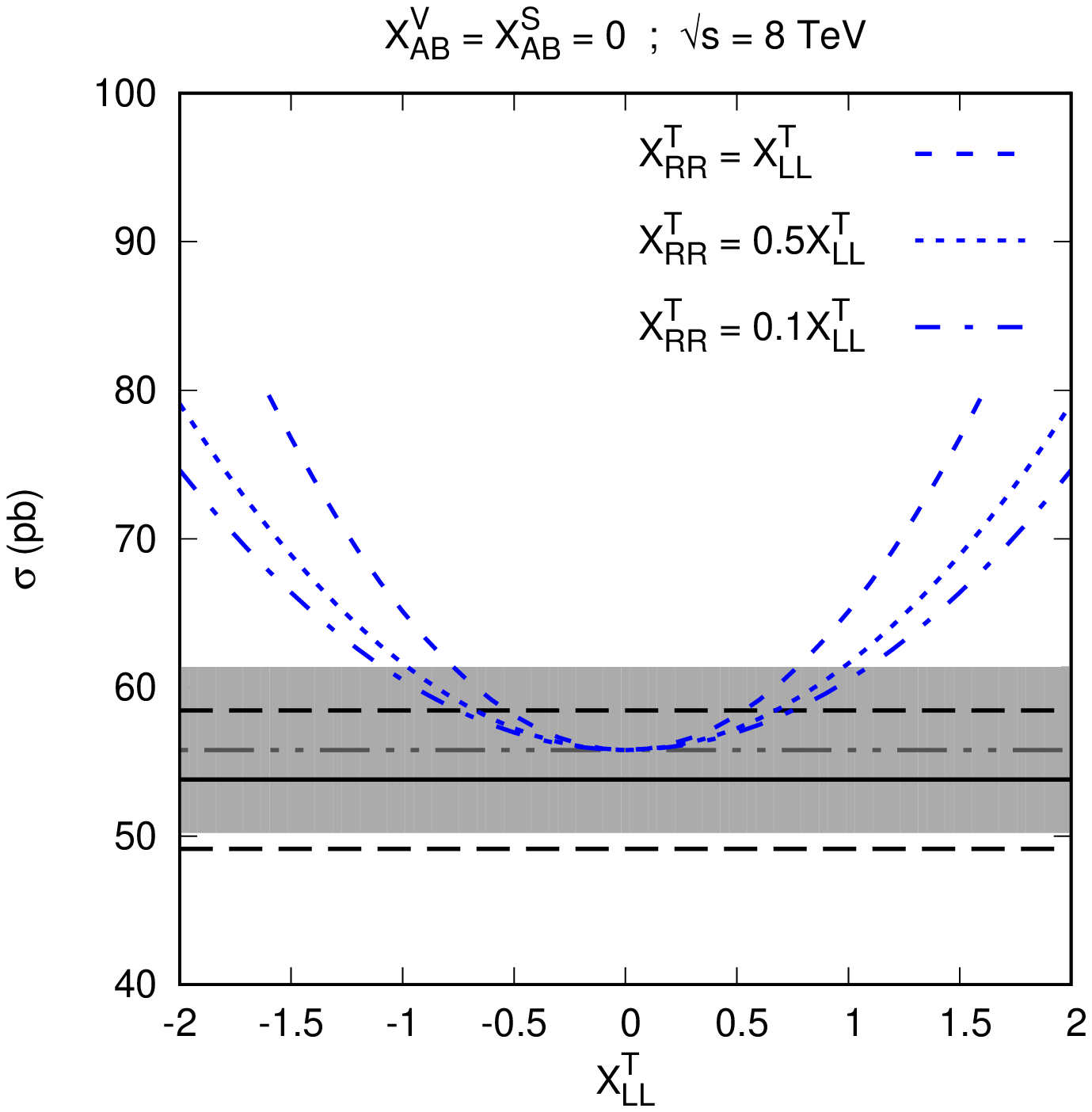}}
\subfigure[]{\includegraphics[scale=0.5]{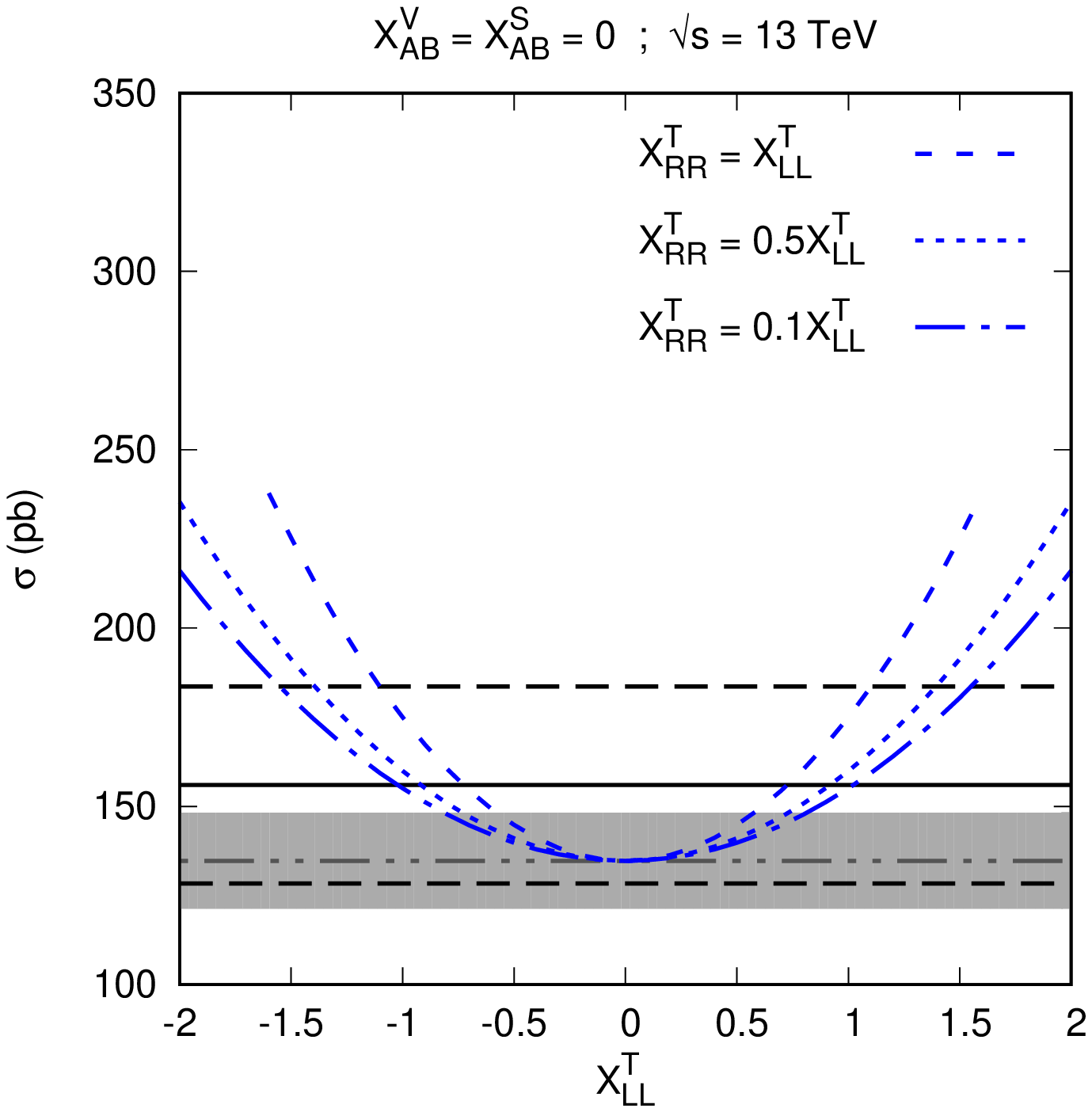}} 
\caption{\reduce $t$-channel single-top production cross section in
  the presence of NP tensor contributions at (a) 8 TeV and (b) 13 TeV.
  The horizontal lines and shaded regions have the same meanings as in
  Fig.~\ref{fig:t-ch_V}.}
\label{fig:t-ch_T}
\end{figure}

In Figs.~\ref{fig:t-ch_8TeV_all} and \ref{fig:t-ch_13TeV_all}, we have
allowed all chiral structures associated with a given Lorentz
structure to have the same weight. That is, when vector operators are
considered ($X^S_{AB} = X^T_{AB} = 0$), we have set $X^V_{LL} =
X^V_{LR} = X^V_{RL} = X^V_{RR}$, and similarly for scalar and tensor
operators. In the following, we relax this condition and consider
scenarios where $X^V_{LR}$, $X^V_{RL}$ and $X^V_{RR}$ are smaller than
$X^V_{LL}$. As expected, this relaxes the constraint on $X^V_{LL}$
(see Fig.~\ref{fig:t-ch_V}). In a UV-complete scenario, these
operators may not all occur simultaneously and there would exist several
possibilities for the relative sizes of the corresponding couplings.
We illustrate the effect using one such possibility. The same exercise
can be carried out for scalar and tensor operators; the results are
depicted in Figs.~\ref{fig:t-ch_S} and \ref{fig:t-ch_T}.

%%%%%%%%%%%%%%%%%%%%%%%%%%%%%%%%%%%%%%%%%%%%%%%%%%%%%%%%%%%%%%%%%%%%%%
\subsection{$s$-channel single-top production}
\label{sec:s_channel}
\begin{figure}[!htbp]
\includegraphics[scale=0.5]{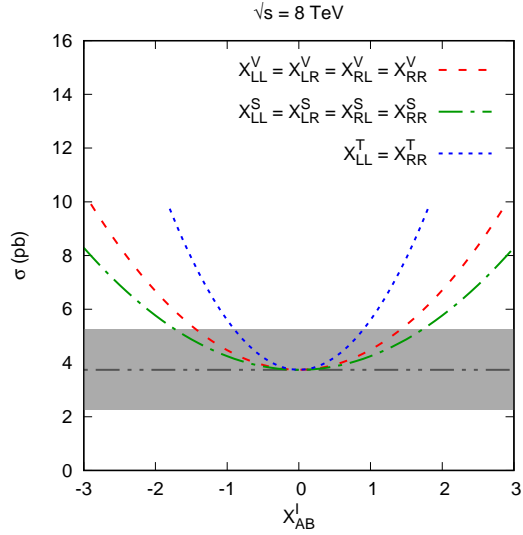}
\caption{\reduce $s$-channel single-top production cross section at
  $\sqrt{s}$ = 8 TeV. The grey dot-dashed line and the grey shaded
  region reflect $\sigma_{SM} \pm 40\%$, which is calculated as
  described in the text.  The red dashed, green dot-dashed and 
  blue dotted curves give the cross section in the presence of NP
  vector, scalar and tensor interactions, respectively. Only one
  Lorentz structure is considered at a time.}
\label{fig:s-ch_8TeV_all}
\end{figure}

%%%%%%%%%%%%%%%%%%%%%%%%%%%%%%%%%%%%%%%%%%%%%%%%%%%%%%%%%%%%%%%%%%%%%%
\vspace*{-10pt}
Single-top production in the $s$-channel is due to the processes $u_i
\bar d_j \to t \bar d_k$, with $d_{j,k} \in \{d,s,b\}$ and $u_i \in \{u,c\}$. Of
these, the largest contribution comes from $u \bar d \to t \bar b$,
since the rest of the processes are CKM suppressed. However, the
introduction of $\Leff$ enhances the contribution due to $c \bar b \to
t \bar b$. This is shown in Fig.~\ref{fig:s-ch_8TeV_all}.

\begin{figure}[!htbp]
\includegraphics[scale=0.5]{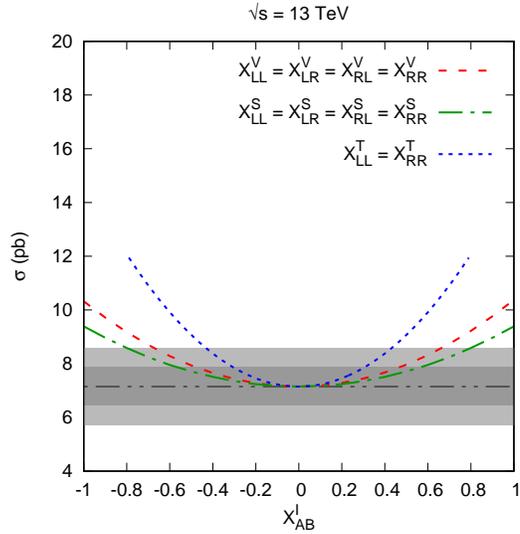}
\caption{\reduce $s$-channel single-top production cross section at
  $\sqrt{s}$ = 13 TeV. The grey dot-dashed line represents
  $\sigma_{SM}$, which is calculated as described in the text. The
  light-grey and dark-grey shaded regions denote the regions
  $\sigma_{SM} \pm 20\%$ and $\sigma_{SM} \pm 10\%$, respectively. The
  red dashed, the green dot-dashed and the blue dotted curves give the
  cross section in the presence of NP vector, scalar and tensor
  interactions, respectively. As noted above, only one Lorentz structure is considered
  at a time.}
\label{fig:s-ch_13TeV_all}
\end{figure}

Since the cross section is small ($\sim$ 3.7 pb in the Standard
Model), CMS~\cite{CMS_s7n8} and ATLAS~\cite{ATLAS_s8} report the
combined cross section due to $p \, p \to t \, X$ and $p \, p \to \bar
t \, X$. However, in our calculations, we consider only $p \, p \to t
\, X$.\footnote{The NP operators considered in this work do, in fact,
  contribute to $p \, p \to \bar t \, X$; we are, however, restricting
  our attention to cases in which a single top quark is produced.}
Hence, in this section, the experimental measurements are not
indicated on the plots. Instead, the grey band denotes a band of
uncertainty around the central value of the corresponding SM cross
section. The size of the band is chosen so as to be commensurate with
the total uncertainty (statistical + systematic) reported in the
actual measurement~\cite{CMS_s7n8,ATLAS_s8}. In Fig.~\ref{fig:s-ch_8TeV_all}, for example, the
size of the band is $\pm$40\%. While measurements in the $s$-channel are 
yet to be reported for 13 TeV, this data set is expected to be several 
times larger than the 8-TeV data set. Hence we use two bands, of sizes 
20\% and 10\%, to project the limits on $X^I_{AB}$ in this case (see Fig.~\ref{fig:s-ch_13TeV_all}).

\begin{figure}[!htbp]
\subfigure[]{\includegraphics[scale=0.5]{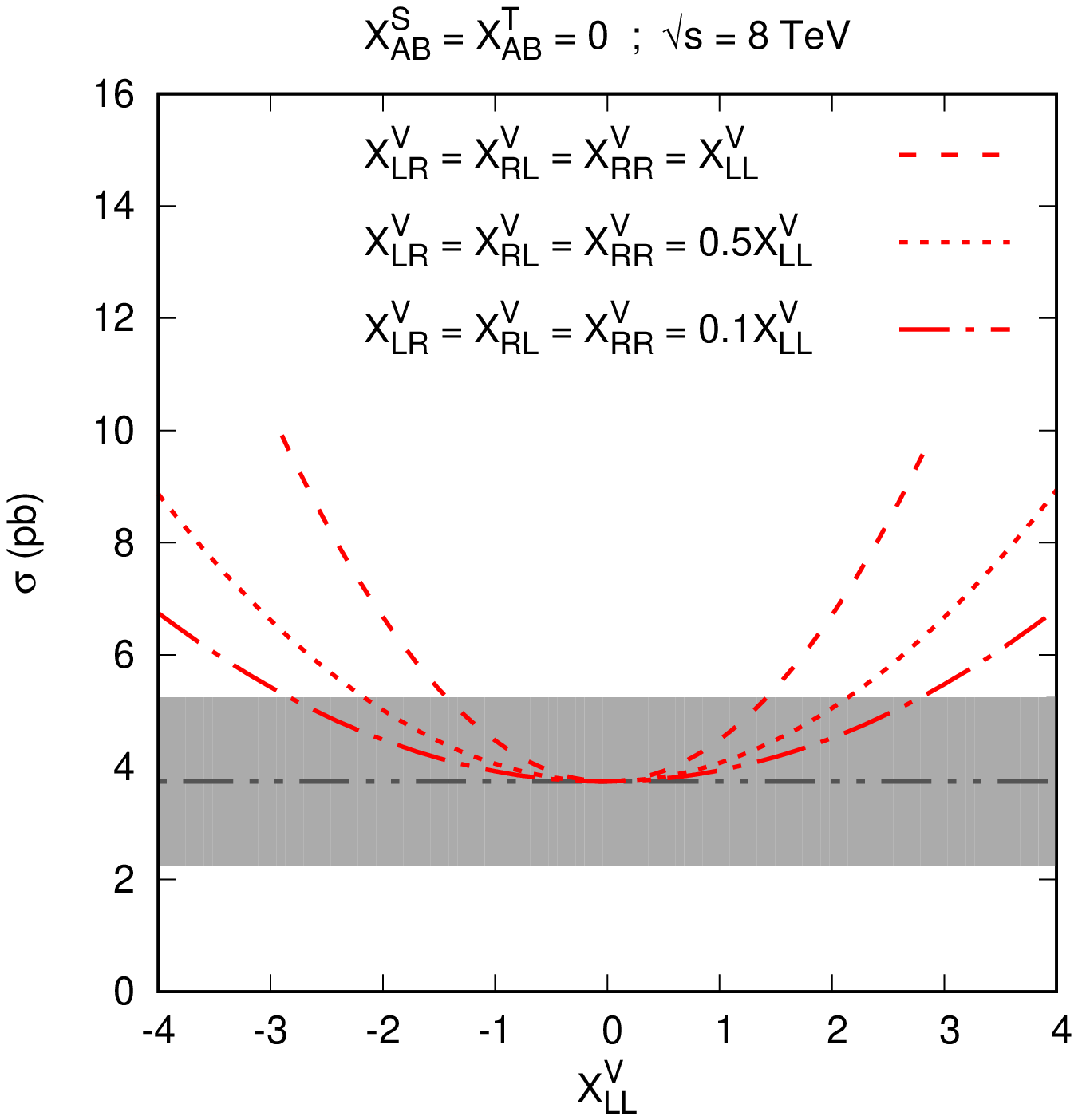}}
\subfigure[]{\includegraphics[scale=0.5]{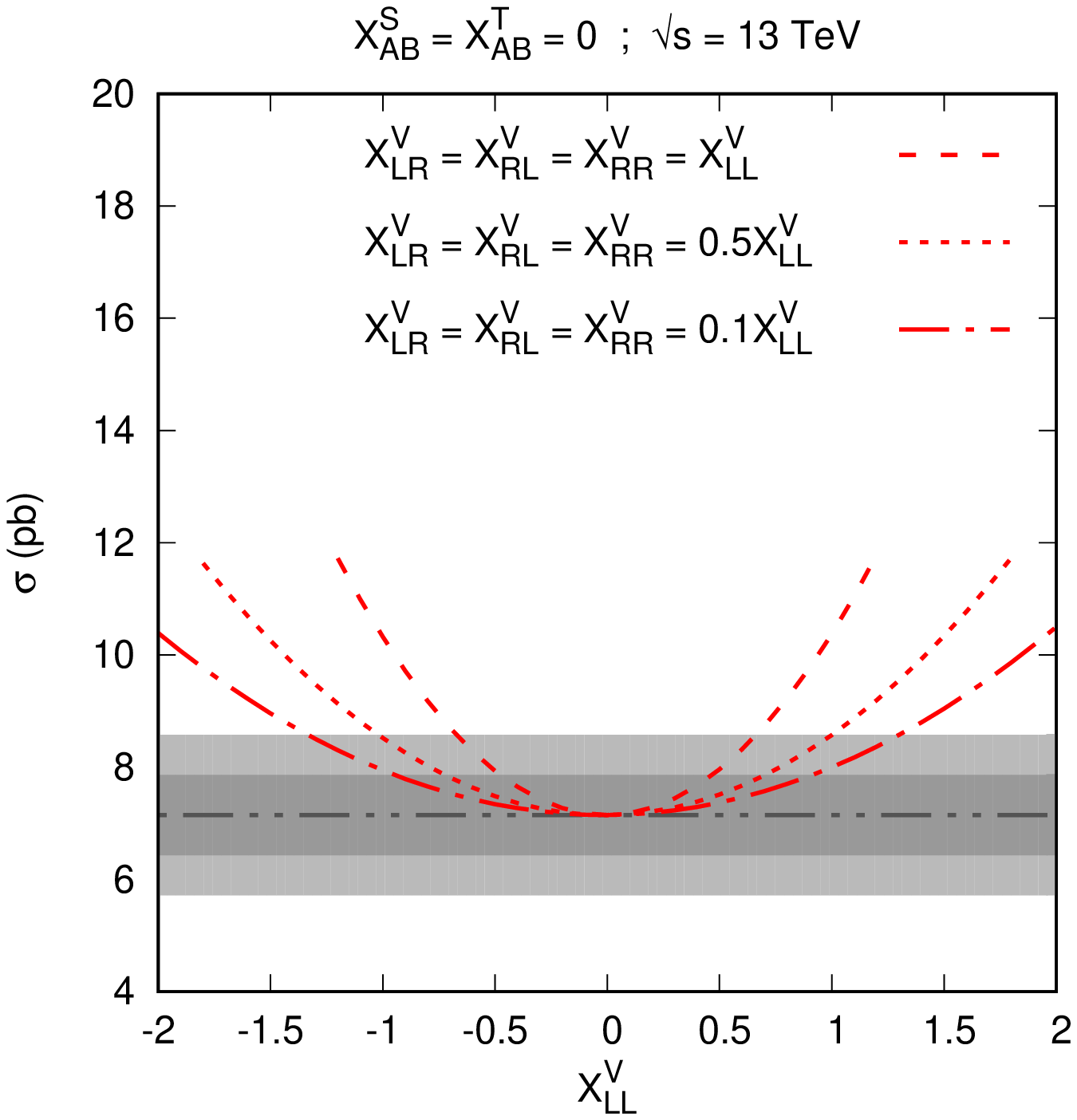}} 
\caption{\reduce $s$-channel single-top production cross section in
  the presence of NP vector interactions at (a) 8 TeV and (b) 13 TeV.  In the
  former case, the grey band depicts $\sigma_{SM} \pm 40\%$; in the
  latter case, the light-grey band depicts $\sigma_{SM} \pm 20\%$ and
  the dark-grey band depicts $\sigma_{SM} \pm 10\%$.}
\label{fig:s-ch_V}
\end{figure}

\begin{figure}[!htbp]
\subfigure[]{\includegraphics[scale=0.5]{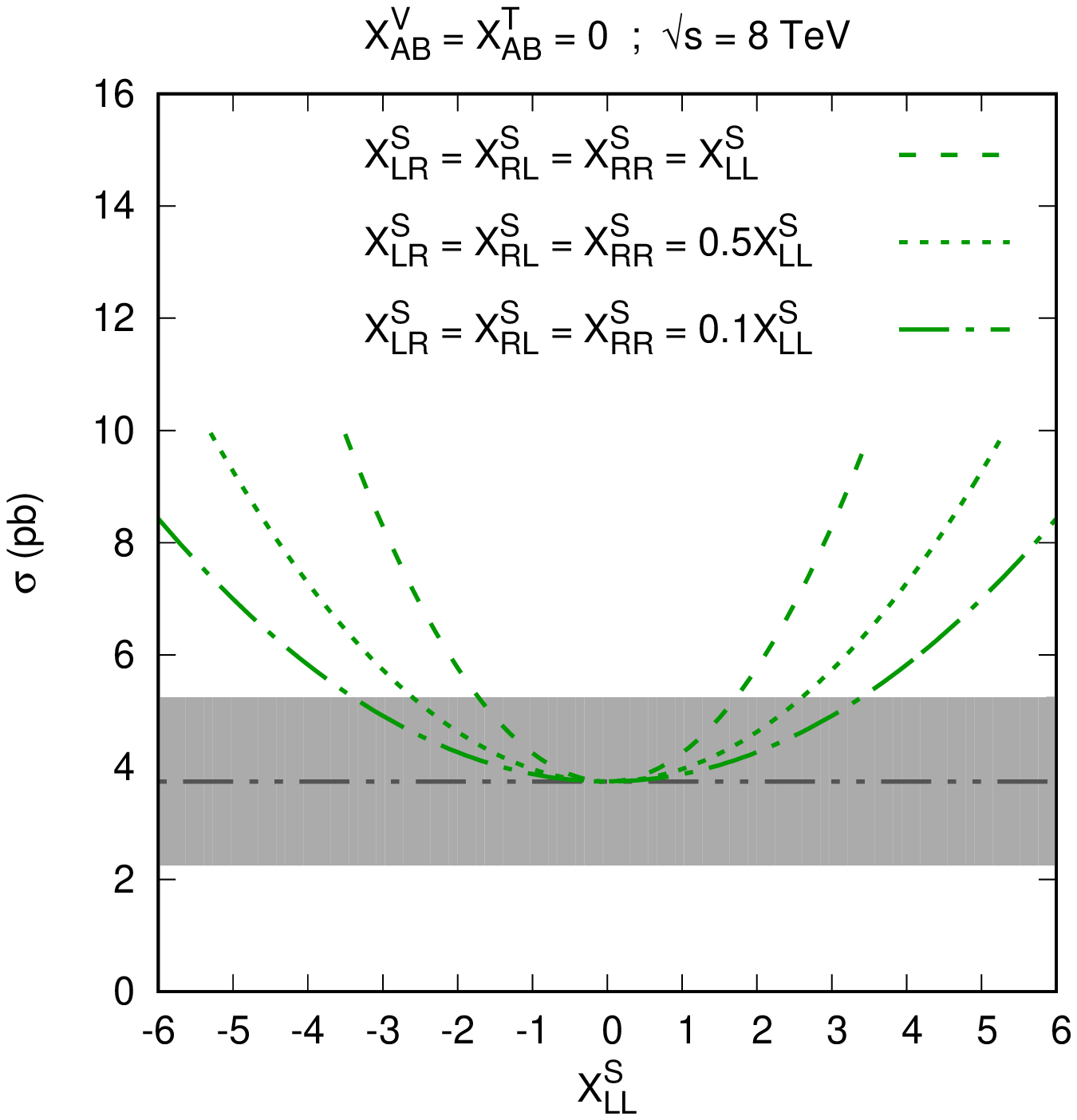}}
\subfigure[]{\includegraphics[scale=0.5]{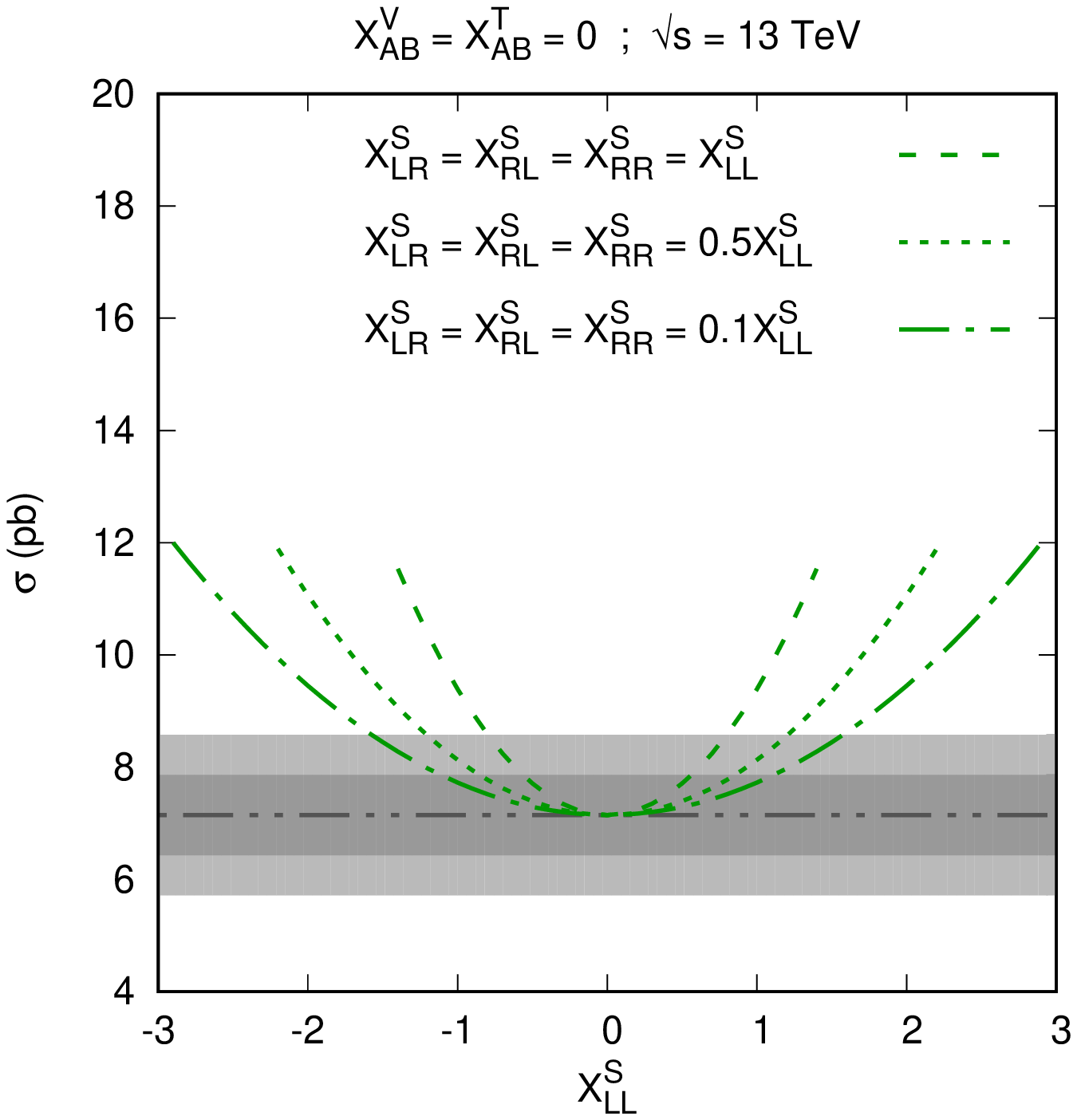}} 
\caption{\reduce $s$-channel single-top production cross section in
  the presence of NP scalar interactions at (a) 8 TeV and (b) 13 TeV.
  The grey bands have the same meanings as in Fig.~\ref{fig:s-ch_V}.}
\label{fig:s-ch_S}
\end{figure}

\begin{figure}[!htbp]
\vspace*{20pt}
\subfigure[]{\includegraphics[scale=0.5]{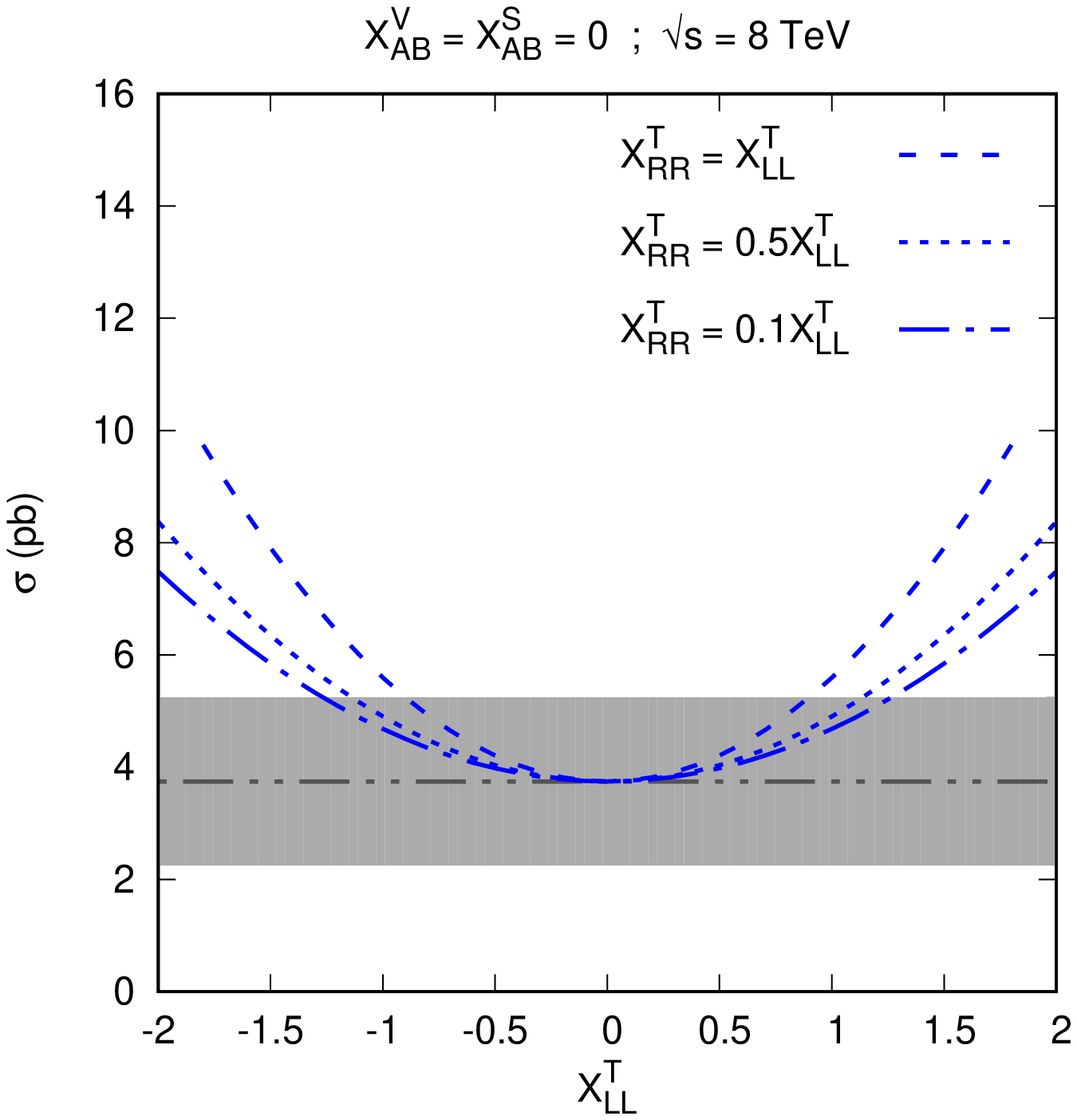}}
\subfigure[]{\includegraphics[scale=0.5]{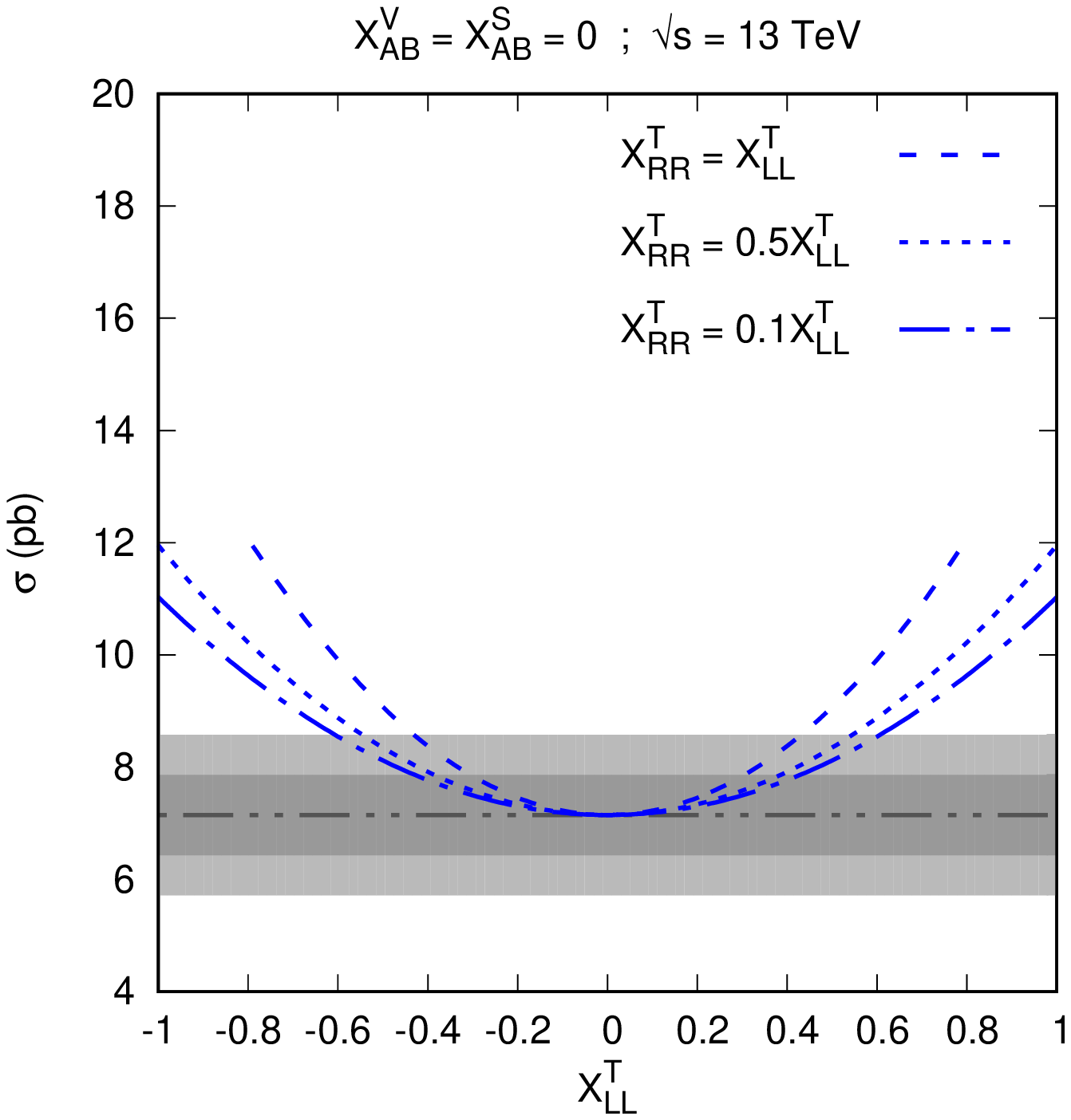}} 
\caption{\reduce $s$-channel single-top production cross section in
  the presence of NP tensor interactions at (a) 8 TeV and (b) 13 TeV.
  The grey bands have the same meanings as in Fig.~\ref{fig:s-ch_V}.}
\label{fig:s-ch_T}
\end{figure}

Comparing Fig.~\ref{fig:s-ch_8TeV_all} with
Fig.~\ref{fig:t-ch_8TeV_all}, it is immediately clear that the
$s$-channel process, even when measured with a lower accuracy, yields
more stringent bounds than the $t$-channel process. This is easy to understand. $\sigma_{SM}$ is
much smaller for the $s$-channel process than for the $t$-channel
process, so even if the relative uncertainty is larger, the absolute
deviation allowed is smaller, which leads to tighter constraints on the couplings.  
A further improvement can be expected to emerge from the $s$-channel
measurements at 13 TeV (see Fig.~\ref{fig:s-ch_13TeV_all}). Here again we
consider scenarios where one chiral structure is dominant. The impacts
on the corresponding limits are shown in Figs.~\ref{fig:s-ch_V},
\ref{fig:s-ch_S} and \ref{fig:s-ch_T}.

%%%%%%%%%%%%%%%%%%%%%%%%%%%%%%%%%%%%%%%%%%%%%%%%%%%%%%%%%%%%%%%%%%%%%%
\subsection{Differential Distributions}
\label{sec:diff_dist}
%%%%%%%%%%%%%%%%%%%%%%%%%%%%%%%%%%%%%%%%%%%%%%%%%%%%%%%%%%%%%%%%%%%%%%
\vspace*{-10pt}
The ATLAS and CMS collaborations have measured the differential cross section 
in terms of the transverse momentum ($p_T$) and rapidity ($|y|$) of the top quark for 
$t$-channel single-top production at 8 TeV~\cite{ATLAS_t8,CMS_dists_8}. 
CMS has also made a similar measurement at 13 TeV~\cite{CMS_dists_13}.

\begin{figure}[htbp]
\subfigure[]{\includegraphics[scale=0.5]{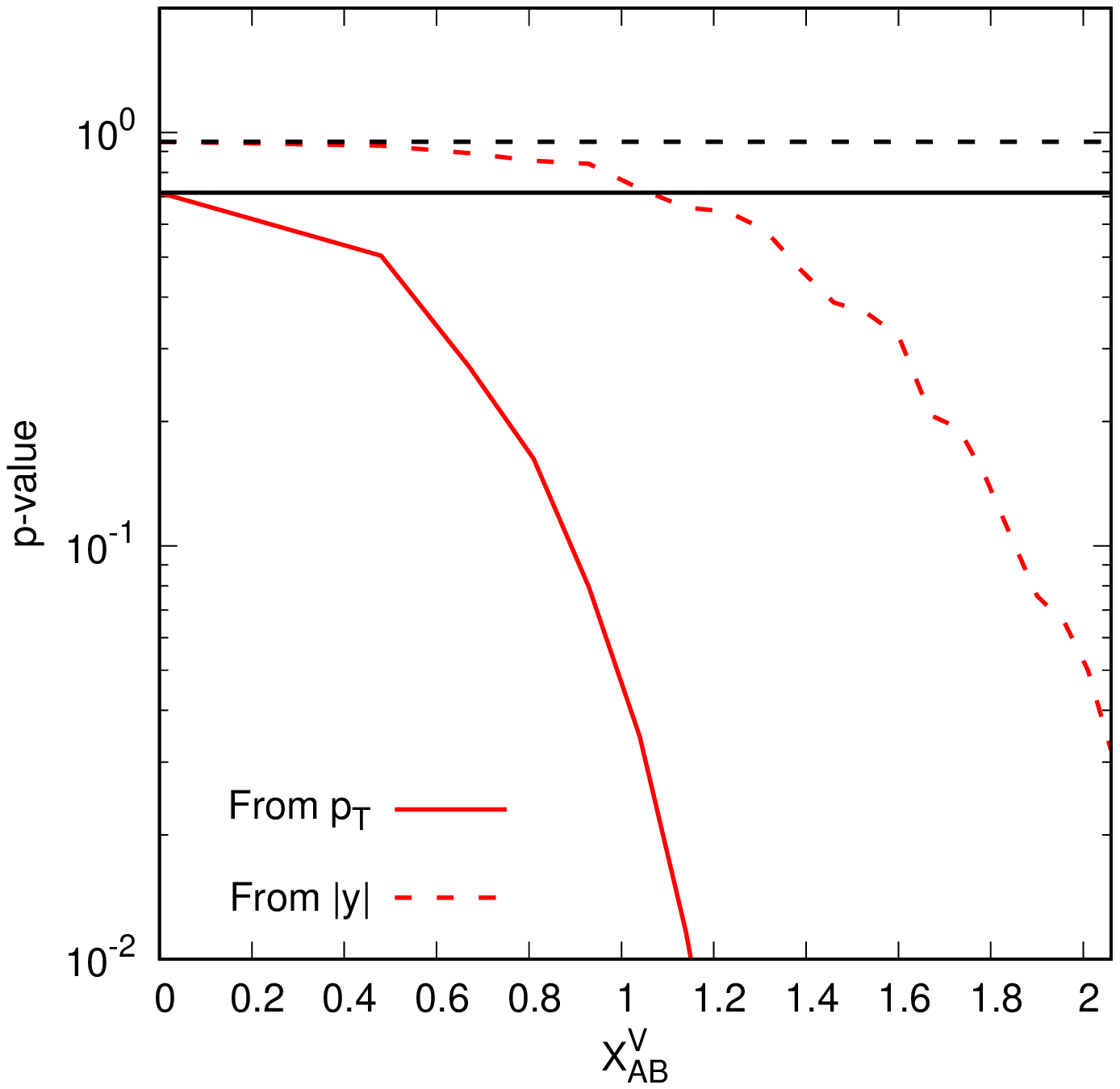}} 
\subfigure[]{\includegraphics[scale=0.5]{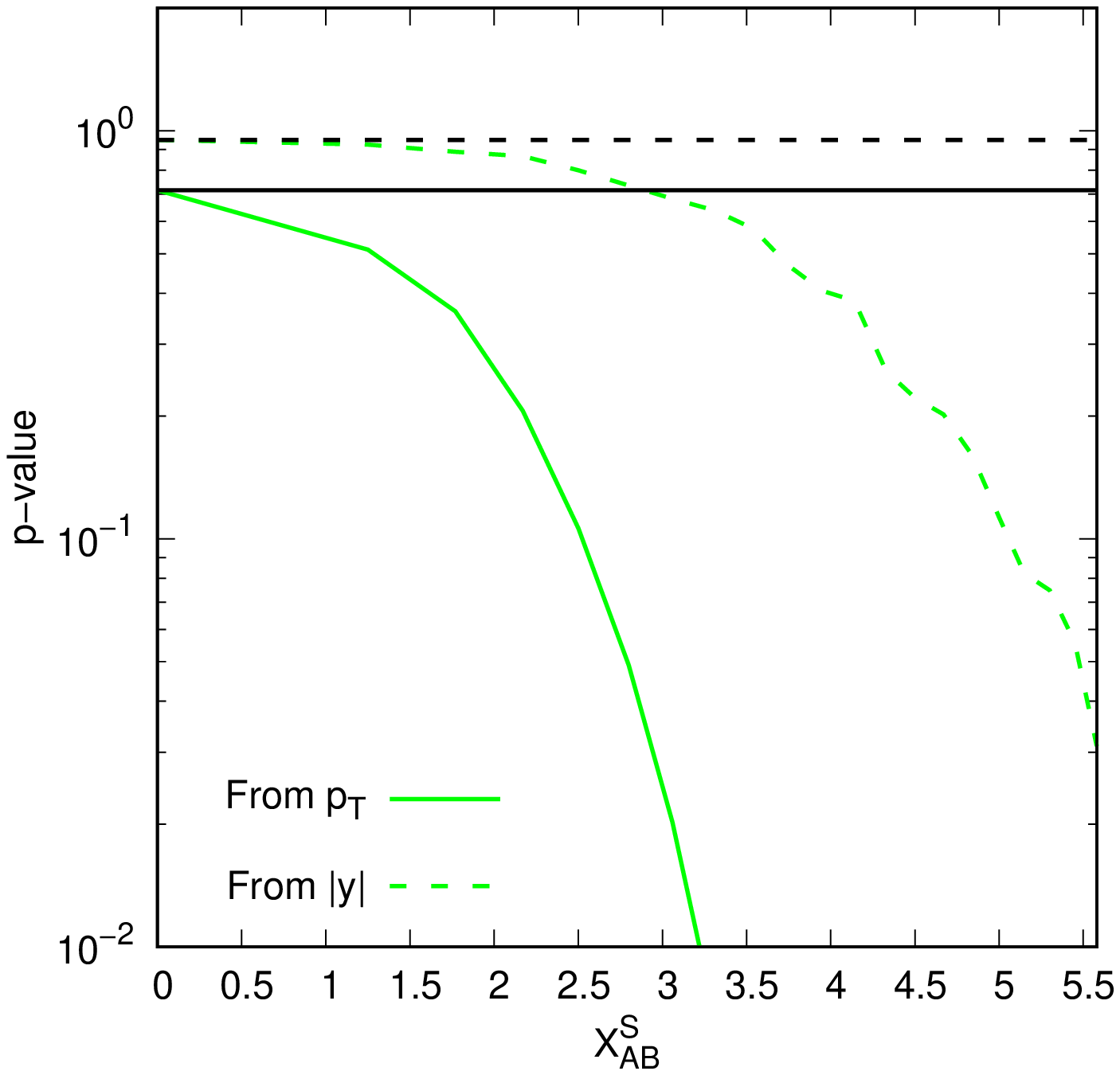}}
\subfigure[]{\includegraphics[scale=0.5]{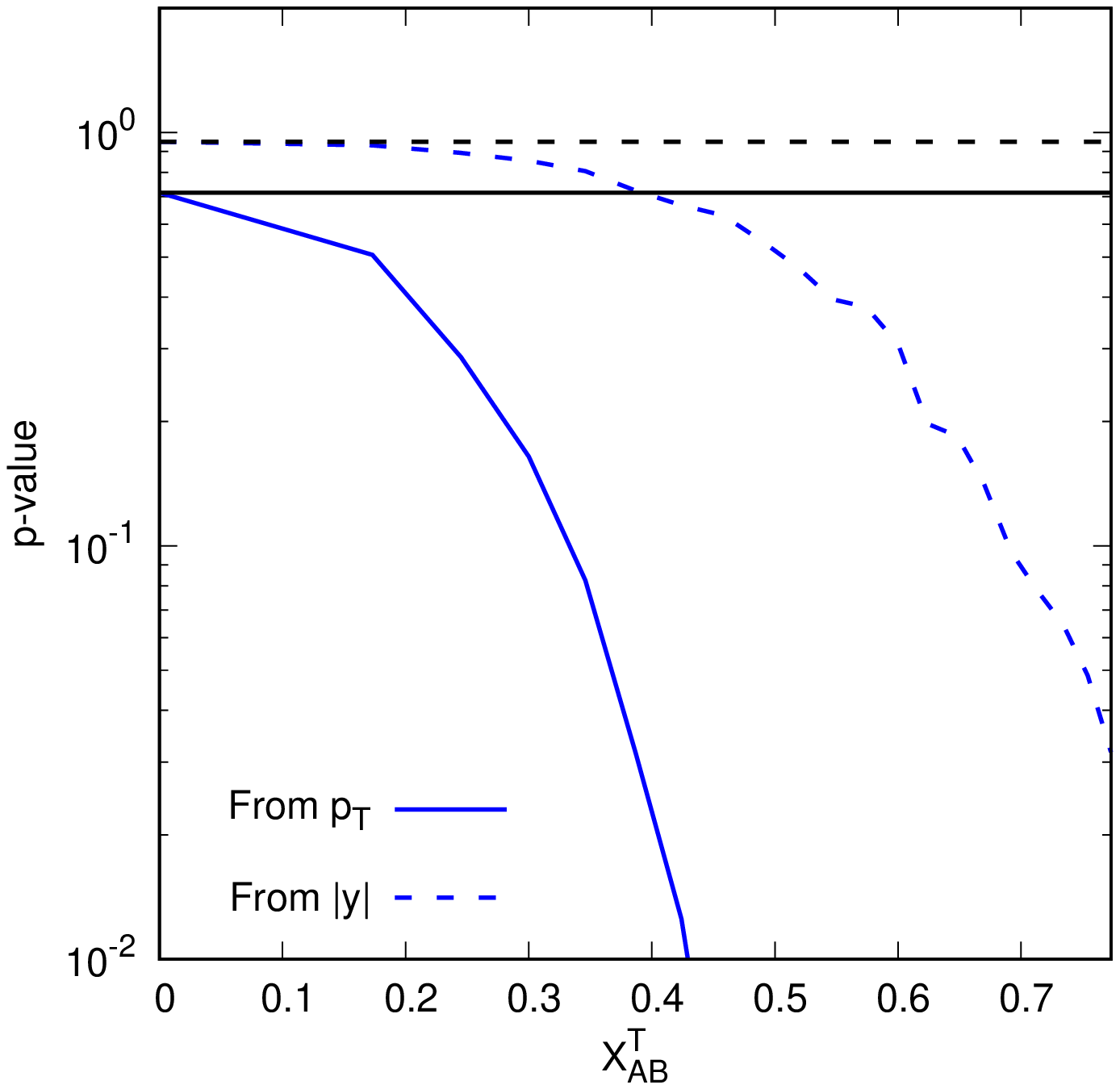}} 
\caption{\reduce $p-$values for various $X^I_{AB}$. Once again we assume 
$X^V_{LL}$ = $X^V_{LR}$ = $X^V_{RL}$ = $X^V_{RR}$ in (a) and analogously
in (b) and (c). The solid and dashed lines correspond to the
results following from the $p_T$ and $|y|$ distributions, respectively.
In each case the horizontal line corresponds to the SM result
and the other line to the case in which NP is included.  See Sec.~\ref{sec:diff_dist}
for further details.
}
\label{fig:fitting_all}
\end{figure}

In Section~\ref{sec:t_channel} we identified the range of $X^V_{AB}$, $X^S_{AB}$ and 
$X^T_{AB}$ that yield a cross section within 10\% of the SM prediction. 
We now confront the predictions from these NP scenarios~\footnote{We now use bin-wise
K-factors along with our LO computations to estimate the distributions at NLO. The K-factors are obtained by 
comparing the LO predictions for the SM with the {\sc Powheg-Box+Pythia6} predictions as given by
Ref.~\cite{ATLAS_t8}.} 
with the ATLAS data~\cite{ATLAS_t8}.
In doing so, we first compute the $\chi^2$ by comparing the ATLAS 8-TeV data for the unfolded, parton-level
distributions
(see Figs.~21(a) and 22(a) and Tables~18 and 20 in Ref.~\cite{ATLAS_t8})
with the distributions computed for the various NP scenarios. This $\chi^2$ is then converted to a $p-$value
by the usual formula
\begin{equation}
p \quad = \quad \int_{\chi^2_{obs}}^{\infty} \, f_{\chi^2}(x;n) \, dx \;,
\label{eq:pvalue}
\end{equation}
where $f_{\chi^2}(x;n)$ is the $\chi^2$ p.d.f. and $n$ is the number of degrees of freedom. 
The results are plotted in Fig.~\ref{fig:fitting_all}. The $p_T$ spectrum proves to have more 
discriminating power than the $|y|$ spectrum. Using this it is possible to put further 
constraints on $X^I_{AB}$. For example, if we assume that NP scenarios that yield $p <$ 0.05
are disfavoured, then the resulting restriction on $X^V_{AB}$, $X^S_{AB}$ and $X^T_{AB}$ is shown in 
Fig.~\ref{fig:limits_comparison}.

%%%%%%%%%%%%%%%%%%%%%%%%%%%%%%%%%%%%%%%%%%%%%%%%%%%%%%%%%%%%%%%%%%%%%%
\subsection{Top polarization}
\label{sec:polarization}
%%%%%%%%%%%%%%%%%%%%%%%%%%%%%%%%%%%%%%%%%%%%%%%%%%%%%%%%%%%%%%%%%%%%%%

The polarization of the top quark is an important observable at the
LHC and a few measurements of it have already been
made~\cite{toppol,toppol_single}. The net polarization is usually
defined as
\begin{equation}
A_P \quad = \quad \dfrac{N_{\uparrow} - N_{\downarrow}}{N_{\uparrow} + N_{\downarrow}} 
\label{eq:AP}
\end{equation}
where $N_{\uparrow}$ and $N_{\downarrow}$, respectively, denote the number of top
quarks with spin aligned along or opposite to a chosen
direction. The value of $A_P$ depends on the choice of the reference
direction. The usual choices for this reference direction are the
$z$-axis or the momentum of the top itself.  In the latter case,
$N_{\uparrow}$ and $N_{\downarrow}$ denote the number of top quarks of
different helicities. If the top quark is produced in association with
another particle (as in the case of single-top production), the
momentum of that particle may also be chosen as the reference
direction. The utility of $A_P$ lies in the fact that it is sensitive
to the chiral structure of the coupling at the production
vertex. Fortuitously, due to its large mass, the top quark decays
before hadronizing. This allows the polarization information (which
would otherwise be lost during hadronization) to be gleaned from the
angular distribution of the top quark's decay products. Very often,
the $e$ or $\mu$ coming from the top decay is used for this purpose.

In the preceding sections, we have identified regions of parameter
space that are compatible with various single-top production
cross section measurements.  We now examine the reach
of $A_P$ as a means to distinguish between the different types of
couplings. In particular, we expect $A_P$ to deviate from its Standard
Model value when there is an increase in the fraction of $t_R$ in the ensemble, 
that is, when there are contributions from $X^I_{RL}$ or
$X^I_{RR}$. As can be inferred from the expressions
in the Appendix, cross section measurements are sensitive to various
combinations of {\em sums} of $\Ahat$'s.  By way of contrast, the spin-dependent contributions
to the amplitude squared are dependent on {\em differences} of 
$\Ahat$'s, such as $\hat{A}_{\bbar}^+-\hat{A}_{\bbar}^-$.
For this reason, polarization measurements can yield additional
information regarding the types of NP interactions that contribute
to the various processes under consideration in this work.
In the following analysis, we choose certain values of $X^I_{AB}$ that yield a
cross section within a certain  ``allowed'' range and then compute $A_P$ according to
Eq.~(\ref{eq:AP}) above, with $N_{\uparrow}$ and $N_{\downarrow}$
denoting the number of top quarks of either helicity.\footnote{That is, the reference axis
    is given by the top quark's momentum.}
Note that we include NP effects in both the numerator and the denominator
when calculating $A_P$.

Table~\ref{tab:asymm_s-ch_8TeV} lists $s$-channel single-top
polarization asymmetries for various combinations of NP contributions
at $\sqrt{s}$ = 8 TeV. In choosing parameters, we have allowed for an
enhancement of $\sim$20\% in the cross section over the SM prediction,
noting that the experimental uncertainty is close to 40\%. We consider
scenarios for which the dominant contribution comes from one
operator,\footnote{This is also more likely from the point of view of
  a UV-complete model.} keeping the sub-dominant couplings at
approximately one-tenth of the dominant one. We find that the
deviation in $A_P$ can be considerable.  Note that the values of $A_P$
listed in Table~\ref{tab:asymm_s-ch_8TeV} are based on the calculation
of tree-level cross sections, as described at the beginning of
Sec.~\ref{sec:numerical}. $K$-factors cancel out in the calculation of
$A_P$. Estimation of $A_P$ based on higher-order calculations
is beyond the scope of this work.  However, the dominant higher-order
corrections would arise from QCD effects and, as such, they are not
expected to alter $A_P$ significantly.

\begin{table}[!htbp]
\begin{scriptsize}
{
\renewcommand{\arraystretch}{1.2}
\renewcommand{\tabcolsep}{0.5cm}
\begin{tabular}{|c|c|c|c|c|}
\hline
              & Dominant Contribution & No Contribution        & $\sigma$ (pb) & $A_{P}$ \\
\hline
SM            & SM                    & $X^I_{AB}$             & 3.7           & -0.68   \\
VP-1          & $X^V_{LL}$            & $X^S_{AB}$, $X^T_{AB}$ & 4.4           & -0.70   \\
VP-2          & $X^V_{LR}$            & $X^S_{AB}$, $X^T_{AB}$ & 4.5           & -0.70   \\
VP-3          & $X^V_{RL}$            & $X^S_{AB}$, $X^T_{AB}$ & 4.5           & -0.43   \\
VP-4          & $X^V_{RR}$            & $X^S_{AB}$, $X^T_{AB}$ & 4.5           & -0.43   \\
SP-1          & $X^S_{LL}$            & $X^V_{AB}$, $X^T_{AB}$ & 4.5           & -0.73   \\
SP-2          & $X^S_{LR}$            & $X^V_{AB}$, $X^T_{AB}$ & 4.5           & -0.73   \\
SP-3          & $X^S_{RL}$            & $X^V_{AB}$, $X^T_{AB}$ & 4.5           & -0.40   \\
SP-4          & $X^S_{RR}$            & $X^V_{AB}$, $X^T_{AB}$ & 4.5           & -0.41   \\
TP-1          & $X^T_{LL}$            & $X^V_{AB}$, $X^S_{AB}$ & 4.5           & -0.65   \\
TP-4          & $X^T_{RR}$            & $X^V_{AB}$, $X^S_{AB}$ & 4.5           & -0.48   \\
\hline
\end{tabular}
}
\end{scriptsize} 
\caption{\reduce $A_P$ for $s$-channel single-top production at
  $\sqrt{s}$ = 8 TeV.  The deviation of the cross section from its SM
  value is of order $20\%$ in each case.}
\label{tab:asymm_s-ch_8TeV}
\end{table}

We have not attempted to estimate the accuracy with which the
$A_P$ can be measured at the LHC.\footnote{We refrain from making such
  an attempt, since background rejection depends heavily on the specific
  algorithm used, which, in turn, often involves boosted decision
  trees~\cite{toppol_single} and other sophisticated analysis tools
  developed and trained by experimentalists to get the best from their
  respective detectors.}  We note, however, that a recent analysis
from CMS~\cite{toppol_single} was able to make a measurement with
approximately 38\% uncertainty. We also point out that the results
from Ref.~\cite{toppol_single} cannot be used to compare directly with
our results because their choice of reference axis is different from
ours.  Nonetheless, it is clear from Table~\ref{tab:asymm_s-ch_8TeV} that accuracies 
\mbox{$\sim$ 10\%} or better
would be needed in order for $A_P$ to be a useful discriminator.

Finally, in Table~\ref{tab:asymm_s-ch_13TeV}, we list the polarization
asymmetry for $s$-channel single-top production at $\sqrt{s}$ = 13
TeV. This time we allow for $\sim$10\% enhancement in the
cross section over the SM prediction. Once again the largest
deviations correspond to contributions from $X^I_{RL}$ and $X^I_{RR}$,
although the net size of the deviation is smaller. This is largely an
artefact of the restriction placed on the overall increase in
cross section; allowing for a larger variation in the cross section
would, in general, allow for greater deviation in $A_P$. It would not,
however, \textit{guarantee} a greater deviation in $A_P$, since the
size of $A_P$ depends on the chiral structure of the dominant
operator.

\begin{table}[!htbp]
\begin{scriptsize}
{
\renewcommand{\arraystretch}{1.2}
\renewcommand{\tabcolsep}{0.5cm}
\begin{tabular}{|c|c|c|c|c|}
\hline
              & Dominant Contribution & No Contribution        & $\sigma$ (pb) & $A_{P}$ \\
\hline
SM            & SM                    & $X^I_{AB}$             & 7.1           & -0.67   \\
VP-1          & $X^V_{LL}$            & $X^S_{AB}$, $X^T_{AB}$ & 7.8           & -0.69   \\
VP-2          & $X^V_{LR}$            & $X^S_{AB}$, $X^T_{AB}$ & 7.8           & -0.69   \\
VP-3          & $X^V_{RL}$            & $X^S_{AB}$, $X^T_{AB}$ & 7.8           & -0.55   \\
VP-4          & $X^V_{RR}$            & $X^S_{AB}$, $X^T_{AB}$ & 7.8           & -0.53   \\
SP-1          & $X^S_{LL}$            & $X^V_{AB}$, $X^T_{AB}$ & 7.8           & -0.71   \\
SP-2          & $X^S_{LR}$            & $X^V_{AB}$, $X^T_{AB}$ & 7.8           & -0.71   \\
SP-3          & $X^S_{RL}$            & $X^V_{AB}$, $X^T_{AB}$ & 7.8           & -0.54   \\
SP-4          & $X^S_{RR}$            & $X^V_{AB}$, $X^T_{AB}$ & 7.8           & -0.52   \\
TP-1          & $X^T_{LL}$            & $X^V_{AB}$, $X^S_{AB}$ & 7.9           & -0.65   \\
TP-4          & $X^T_{RR}$            & $X^V_{AB}$, $X^S_{AB}$ & 7.9           & -0.56   \\
\hline
\end{tabular}
}
\end{scriptsize} 
\caption{\reduce $A_P$ for $s$-channel single-top production at
  $\sqrt{s}$ = 13 TeV.  The deviation of the cross section from its SM
  value is of order $10\%$ in each case.}
\label{tab:asymm_s-ch_13TeV}
\end{table}

One can, similarly, obtain the polarization asymmetry
for $t$-channel single-top production. If we follow our earlier
strategy of choosing benchmark points from well within the allowed
band, we would be considering scenarios where $\Delta\sigma \sim
0.05 \, \sigma_{SM}$. In that case, the deviation in $A_P$ is
$\lesssim$ 10\% and may not be experimentally discernible. However,
as one allows for larger $\Delta\sigma$, deviations in $A_P$ also
begin to increase. Tables similar to Tables~\ref{tab:asymm_s-ch_8TeV} 
and \ref{tab:asymm_s-ch_13TeV} for the $t$-channel are not presented 
here for the sake of brevity.

%%%%%%%%%%%%%%%%%%%%%%%%%%%%%%%%%%%%%%%%%%%%%%%%%%%%%%%%%%%%%%%%%%%%%%
\subsection{$t$-channel single-top production - a Futuristic Analysis}
\label{sec:futuristic}
%%%%%%%%%%%%%%%%%%%%%%%%%%%%%%%%%%%%%%%%%%%%%%%%%%%%%%%%%%%%%%%%%%%%%%
\vspace*{-10pt}
We now return to the futuristic analysis that we briefly alluded to in 
the Introduction. As discussed in Sec.~\ref{sec:t_channel}, 
$t$-channel single-top production gets
contributions from all processes of the type $d_i u_j \to t d_k$ and
$d_i \bar d_j \to t \bar u_k$ where $u_{j,k} \in \{u,c\}$ and $d_{i,j,k} \in \{d,s,b\}$.  
The new physics operators, however, only affect the
sub-processes $b c \to t b$ and $b \bar b \to t \bar c$.  If one of
these sub-processes could be isolated and studied separately, then it
would be possible to obtain far more stringent constraints on the new
physics parameters. The reason for the enormous increase in
sensitivity is obvious -- in the earlier cases, the NP
contribution arising from just two subprocesses had to compete with
the SM background arising from a multitude of
subprocesses. In this case, it would be competing with background
arising from just one subprocess. This also resolves the conundrum
that we encountered in Sec.\ref{sec:t_channel} -- if $X^I_{AB}$ of 
$\cal O$(1) (or, equivalently $\Lnew$ $\lesssim$ 1 TeV) are not ruled out, 
why hasn't new physics been discovered at the LHC ? It is because the 
sensitivity of the LHC in this context is limited by the fact that one or two 
NP amplitudes have to compete against a large number of SM amplitudes.  

From the point of view of isolating $t$-channel subprocesses, $b c \to t b$ 
appears at first glance to be more promising than $b \bar b \to t \bar c$, 
since $b c \to t b$ events could be isolated from other
\mbox{$t$-channel} events by the application of an additional
$b$-tag. In reality, the situation is somewhat complicated.

In order to identify $s$-channel single-top production and distinguish
it from $t$-channel single-top production, experiments already use an
additional $b$-tag. The idea here is that if a top quark is produced
in an $s$-channel process, then about 99\% of the time, it is accompanied 
by a bottom quark owing to the strength of $V_{tb}$. In contrast, 
a $tb$ final state is rare in the $t$-channel process.\footnote{A $t d$
final state is approximately $10^4$ times more likely to occur than
a $t b$ final state in $t$-channel single-top production.}  Since
$b$-tags do not distinguish between $b$ and $\bar b$, both $t b$ and
$t \bar b$ final states are identified as coming from the $s$-channel
process. The contamination in $s$-channel measurements arising out of
such misidentifications is not significant, since the cross section for
the ($t$-channel) $tb$ final state is orders of magnitude smaller than
the dominant $s$-channel contribution.

\begin{figure}[htbp]
\vspace*{40pt}
\subfigure[]{\includegraphics[scale=0.5]{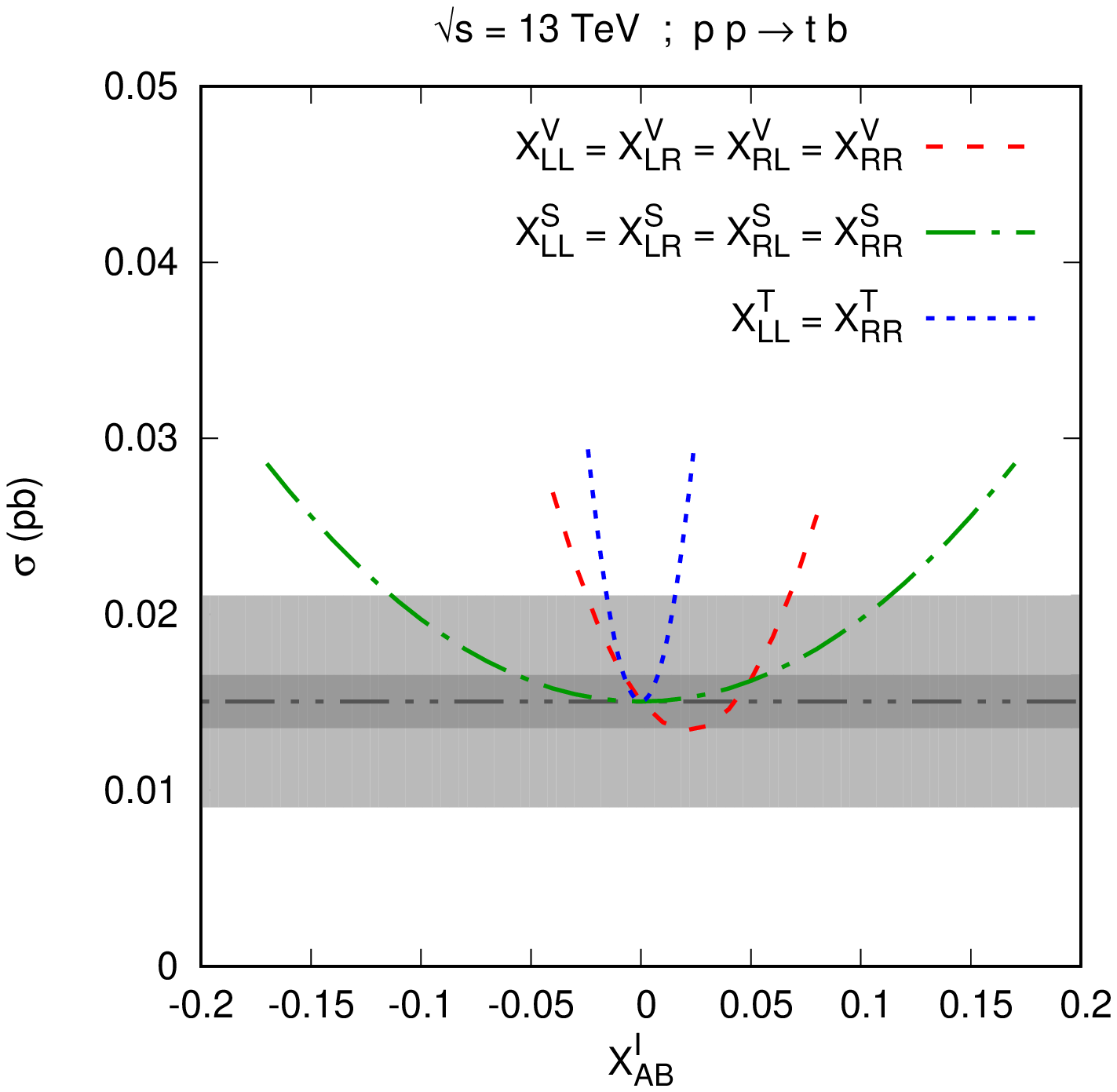}} 
\subfigure[]{\includegraphics[scale=0.5]{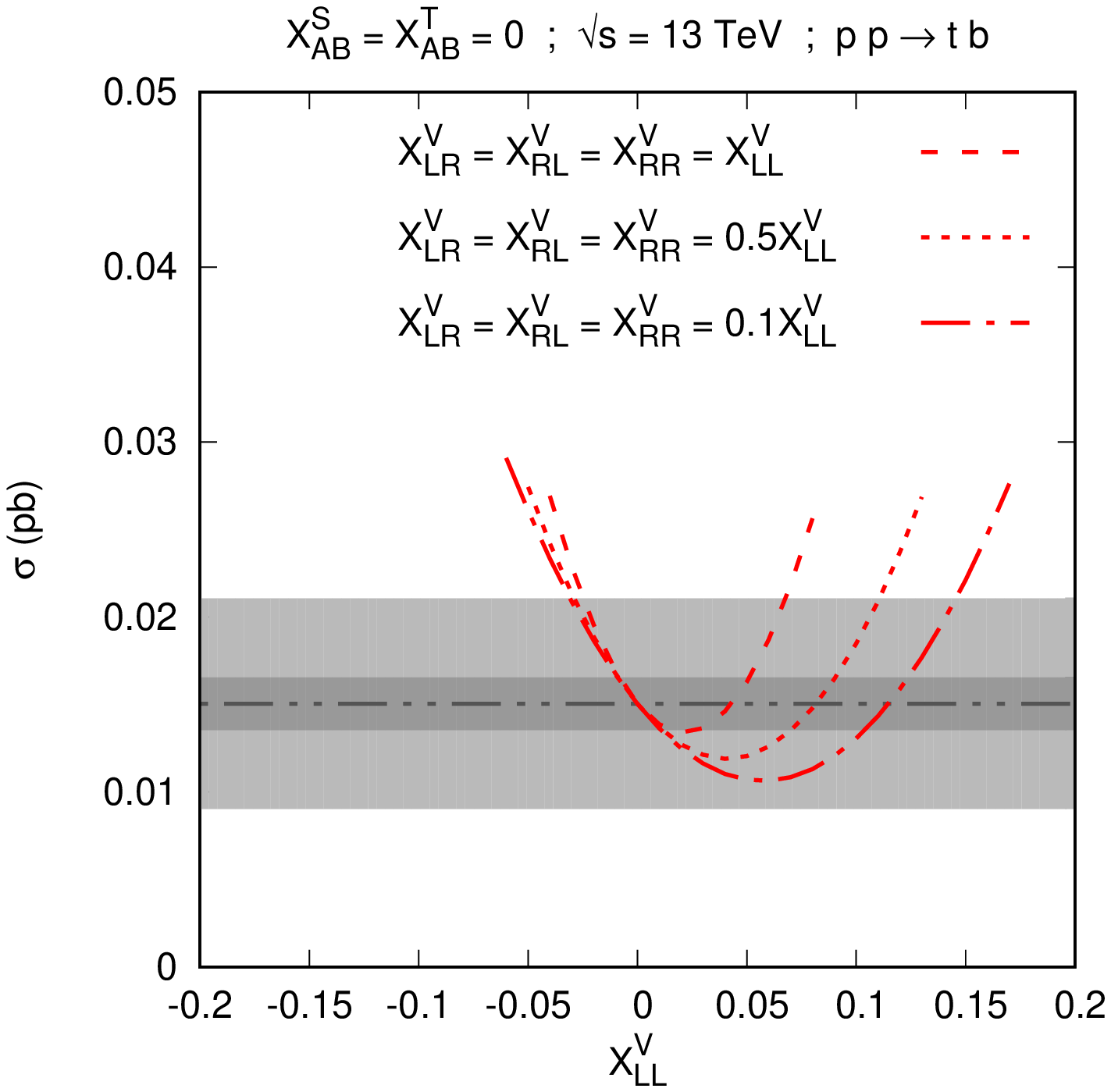}} 
\subfigure[]{\includegraphics[scale=0.5]{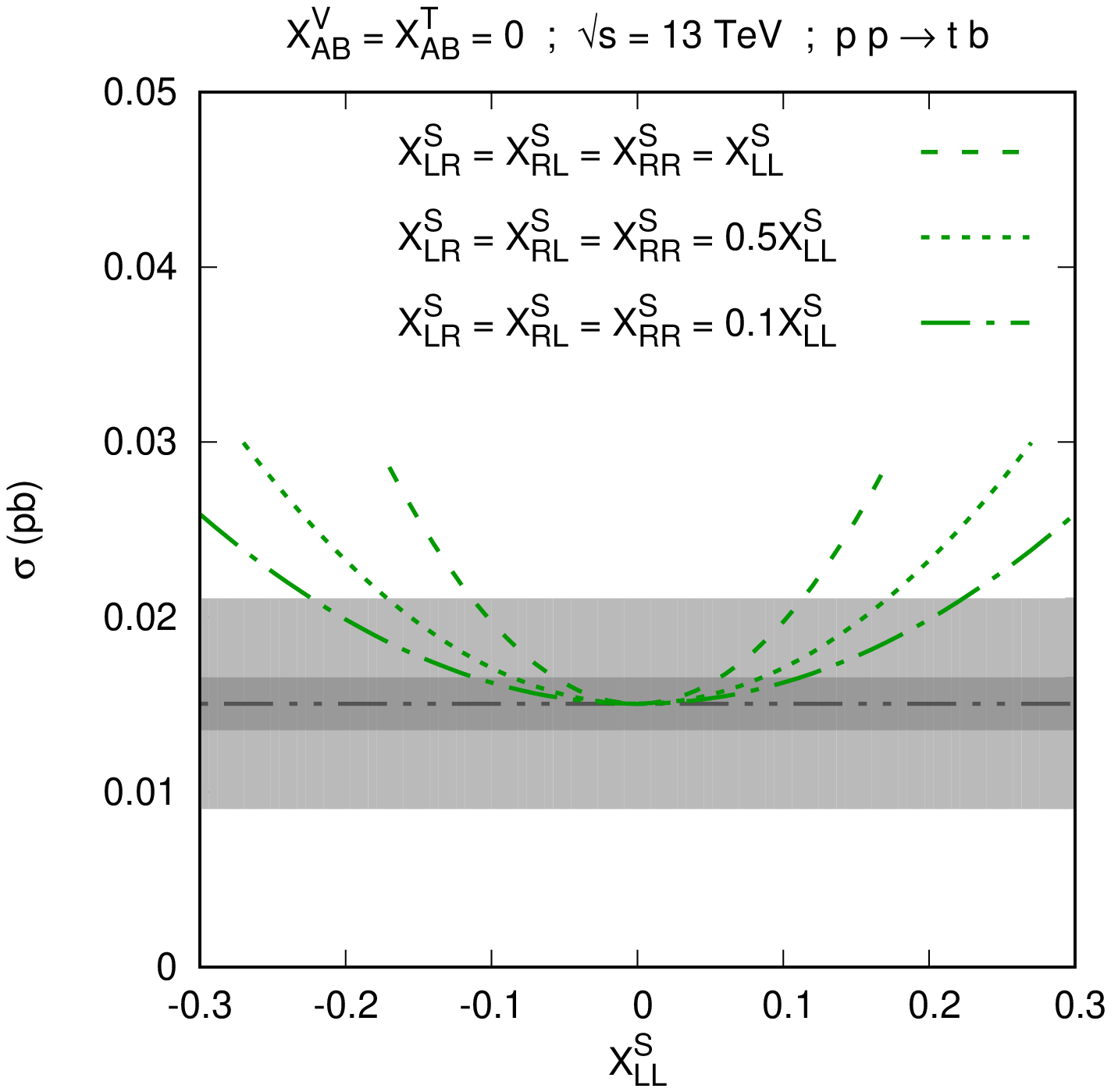}} 
\subfigure[]{\includegraphics[scale=0.5]{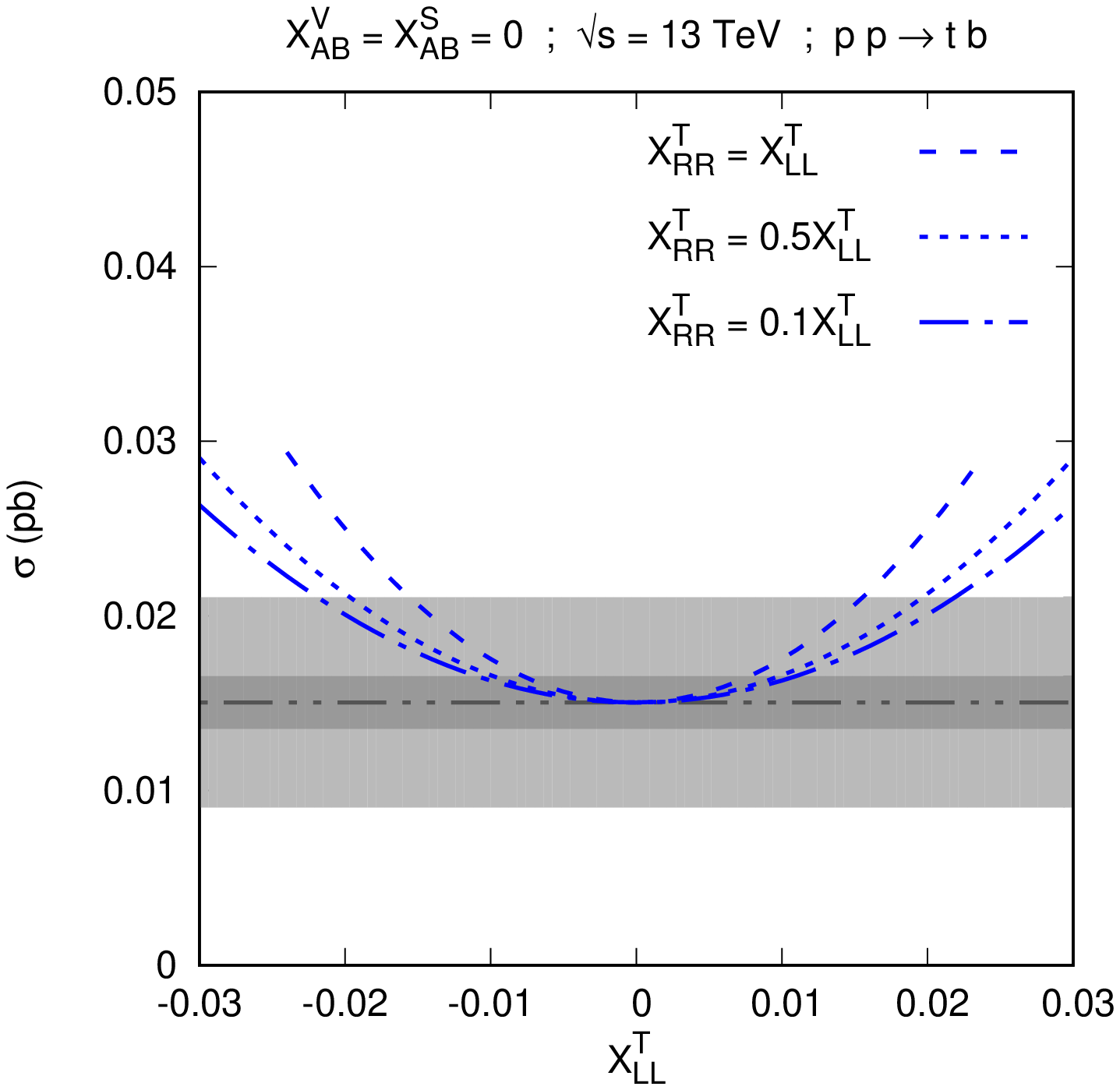}} 
\caption{\reduce Single-top production cross section in the channel $pp \to tb$ in the presence of $X^I_{AB}$ at 13 TeV.
The light-grey band depicts $\sigma_{SM} \pm 40\%$, while the dark-grey band depicts $\sigma_{SM} \pm 10\%$.}
\label{fig:t-ch_tb_13TeV_all}
\end{figure}

However, if it were possible to distinguish between $b$ and $\bar b$
quarks, then one would be able to isolate the process $b c \to t b$. 
To estimate the expected improvement in the limits, one only needs
to consider the size of the SM background, which, at 13 TeV, is
approximately 135 pb for the usual $t$-channel production and about
0.014 pb when $b c \to t b$ is isolated. The actual limits are
depicted in Fig.~\ref{fig:t-ch_tb_13TeV_all}, which can be compared
with Fig.~\ref{fig:t-ch_13TeV_all} and with the 13-TeV plots in
Figs.~\ref{fig:t-ch_V}-\ref{fig:t-ch_T}. $\sigma_{SM} \pm 10\%$
is depicted as a grey band in each of these cases. 
In Fig.~\ref{fig:t-ch_tb_13TeV_all}(a)
and (b), the effect of the interference between the SM and the NP
$X^V_{LL}$ term is discernible, unlike in the corresponding plots in
Sec.~\ref{sec:t_channel}.

While isolating $b c \to t b$ seems to be an exciting possibility,
techniques for distinguishing between $b$-flavored quarks and
antiquarks in the final state are still at a nascent stage, although
some developments in this direction have been
reported~\cite{bcharge}. If such techniques can be improved upon
sufficiently so as to become reliable even when the statistics are
relatively low, then the $p p \to t b$ channel will become the primary
channel of interest.

%%%%%%%%%%%%%%%%%%%%%%%%%%%%%%%%%%%%%%%%%%%%%%%%%%%%%%%%%%%%%%%%%%%%%%
\section{Summary and Conclusions}
\label{sec:summary}
%%%%%%%%%%%%%%%%%%%%%%%%%%%%%%%%%%%%%%%%%%%%%%%%%%%%%%%%%%%%%%%%%%%%%%

Top quark decays are most sensitive to new physics effects at the
energy scale of a few hundred GeV.  The same effects, if they arise
from a higher energy scale, would be more effectively probed in
single-top production.  In this work, we have focused on NP effects
arising from anomalous couplings between the top, bottom and charm
quarks.  Since we only consider the top's interactions with
heavy quarks, it might at first appear that progress would be thwarted
by the low densities of heavy quarks inside the proton. Nevertheless, our
detailed study shows that it would be possible to place meaningful
constraints on the new physics parameters.

We have considered $t$-channel and $s$-channel single-top production
cross sections to obtain constraints on contact interactions involving
$t$, $b$ and $c$ quarks. Of these two channels, the stronger
constraints arise from the $s$-channel. This is due to the fact that
the Standard Model background cross section is smaller for the
$s$-channel. Within a given channel ($t$ or $s$), the limits are most
stringent for tensor operators, followed by vector and scalar
operators, respectively. This is essentially due to the additional
numerical factors that appear from the Dirac traces for each of these
operators.

For $t$-channel single-top production, data is also 
available for the $p_T$ and $|y|$ differential cross sections. We have 
examined the compatibility of these measurements with different NP 
scenarios and found that the $p_T$ distribution in particular can be useful 
in further constraining the NP parameter space.

Apart from the total and differential cross sections,
we have also considered the
relative contributions to the cross section from top quarks of
different helicities. The polarization asymmetry $A_P$, which compares
the helicity states of the top quark, can be particularly useful in
establishing the presence of operators involving $t_R$, especially
since the corresponding Standard Model charged current coupling
involves only $t_L$.

\begin{figure}[!htbp]
\subfigure[]{\includegraphics[scale=0.44]{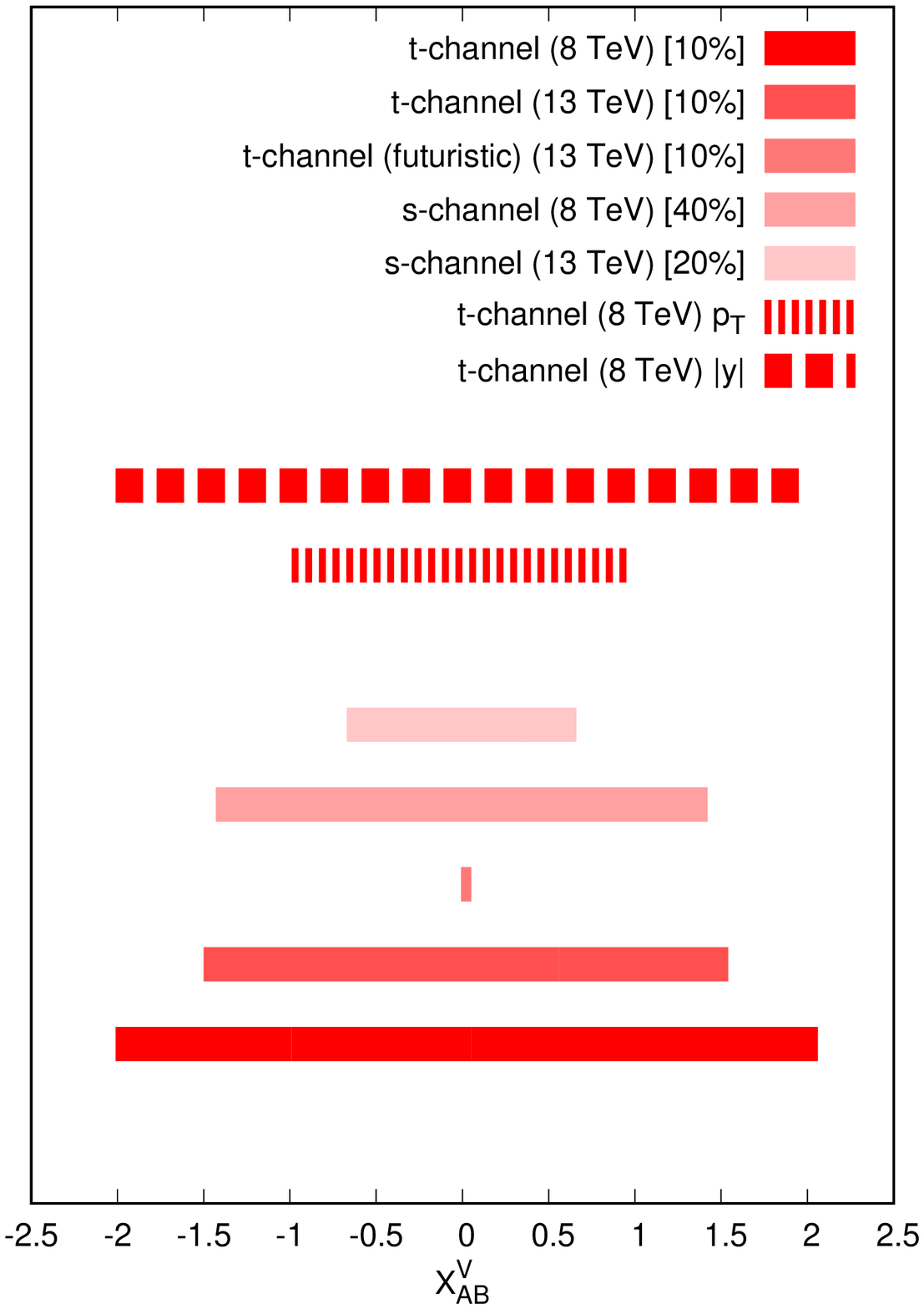}}
\subfigure[]{\includegraphics[scale=0.44]{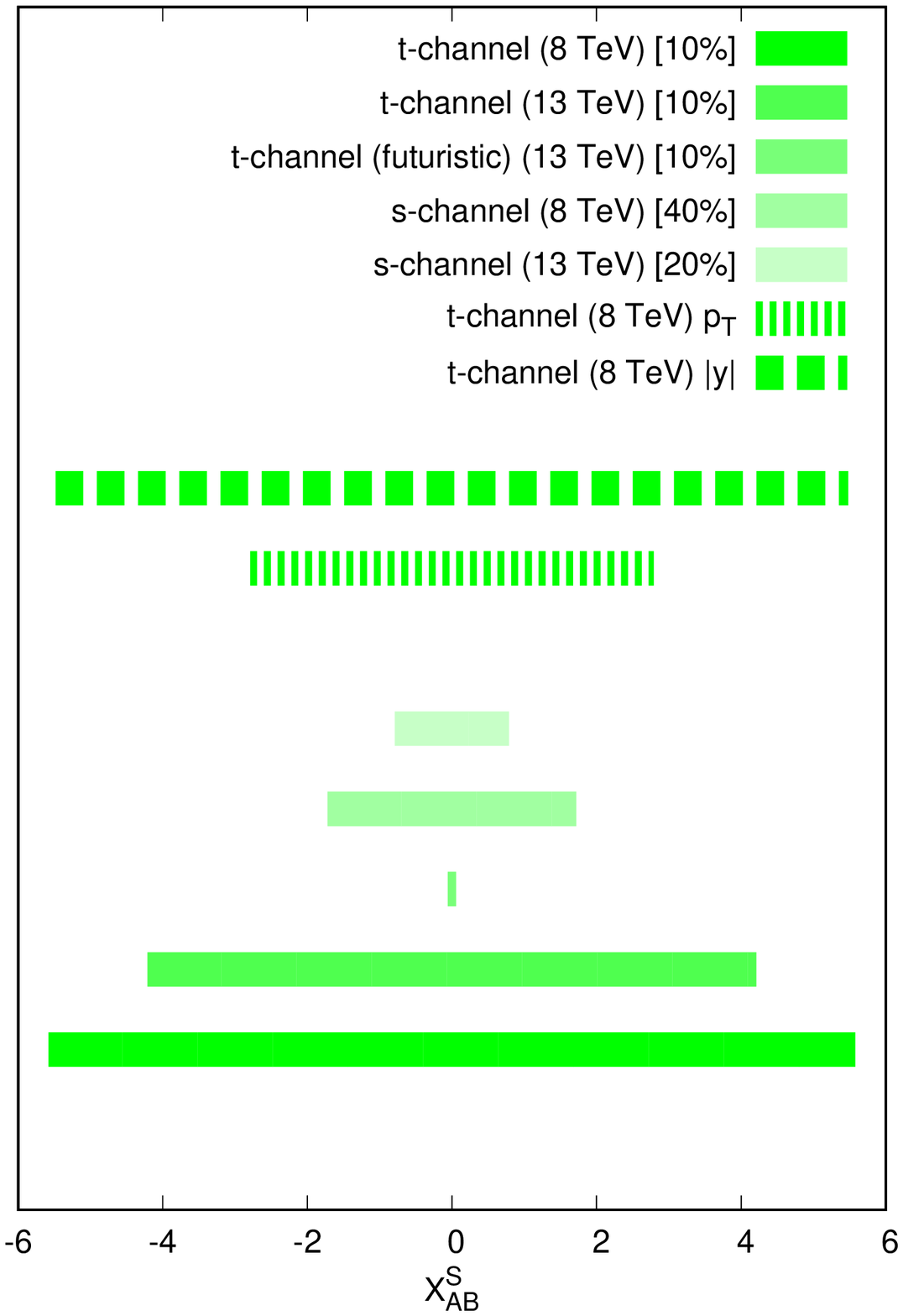}} 
\subfigure[]{\includegraphics[scale=0.44]{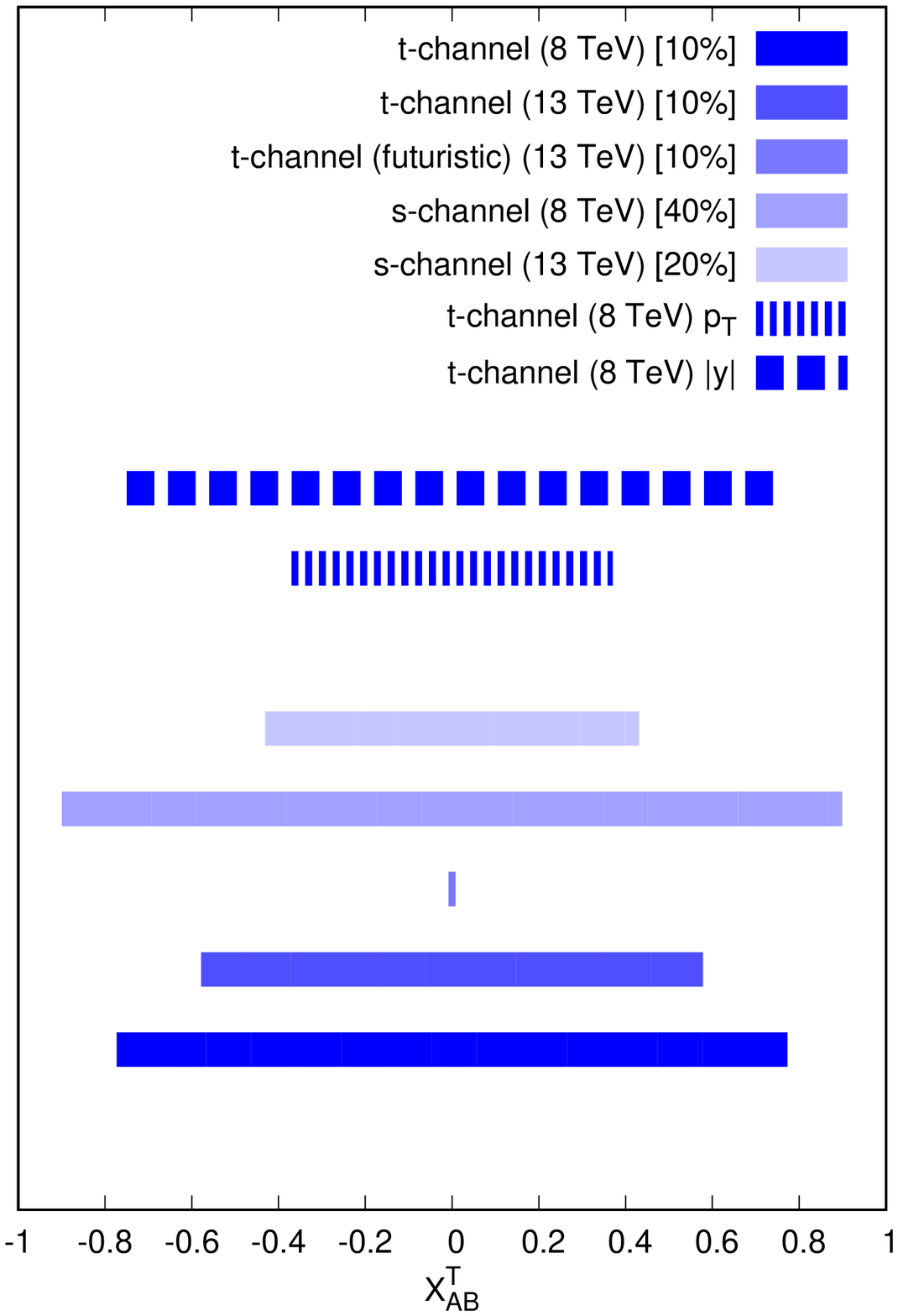}} 
\caption{\reduce Comparison between the limits obtained on
  (a) $X^V_{AB}$, (b) $X^S_{AB}$ and (c) $X^T_{AB}$ from the different
  channels. The numbers in square brackets indicate the allowed
  deviation from the SM cross section, \textit{e.g.} 10\%
  indicates $\Delta\sigma = 0.1\sigma_{SM}$. 
  For (a) we set $X^V_{LL}$ = $X^V_{LR}$ = $X^V_{RL}$ = $X^V_{RR}$. Similarly
  for (b) and (c). 
  Also shown are the limits obtained by imposing the condition $p >$ 0.05
  on the $p_T$ and $|y|$ distributions.
  For details, see Secs.~\ref{sec:t_channel}, \ref{sec:s_channel}, \ref{sec:diff_dist} and
  \ref{sec:futuristic}.}
\label{fig:limits_comparison}
\end{figure}

Finally, we have considered a futuristic analysis. This analysis rests
on one crucial assumption, namely that $b$ quarks in the final state
can be distinguished from $\bar b$ antiquarks on an event-by-event
basis. Adopting this assumption, we have obtained limits on the NP
four-quark operators that are far more stringent than those obtained
from the regular $t$-channel and $s$-channel analyses. The comparison
can be seen in Fig.~\ref{fig:limits_comparison}. The improvement is
indeed startling and perhaps adequate motivation for the pursuit of
experimental techniques that will make it possible.

% \newpage

\begin{center}
\begin{small}\textbf{ACKNOWLEDGEMENTS}\end{small}
\end{center}

\noindent
The authors wish to thank David Cusick and Kristian Stephens for collaboration 
at early stages of this work, and David London and Bhubanjyoti Bhattacharya for 
helpful discussions. We are most grateful to the ATLAS Collaboration for acceding to our 
request for additional information. We would also like to thank the
referee for pointing out additional experimental results relevant to this analysis.  
This work has been partially supported by CONICET and ANPCyT projects PICT-2013-2266 and PICT-2016-0164 (A.S.).
K.K. was supported by the U.S. National Science Foundation under Grant PHY-1215785.
The work of P.S. was supported by the Dept. of Science and Technology, India under 
the INSPIRE Faculty Scheme Grant IFA14-PH-105.

% \vspace*{30pt}

%%%%%%%%%%%%%%%%%%%%%%%%%%%%%%%%%%%%%%%%%%%%%%%%%%%%%%%%%%%%%%%%%%%%%%
\begin{center}
\section*{APPENDIX}       
\end{center}
%%%%%%%%%%%%%%%%%%%%%%%%%%%%%%%%%%%%%%%%%%%%%%%%%%%%%%%%%%%%%%%%%%%%%%
\begin{appendices}

\vspace*{-10pt}

\renewcommand\thesubsection{\arabic{subsection}}
\renewcommand{\theequation}{\Alph{section}.\arabic{equation}}

In this appendix we write down parton-level expressions for the matrix
element squared for the $s$- and $t$-channel processes considered in
this paper.  Throughout, we use the notation $Q^2 = (p_1 + p_2)^2$ and
$q^2 = (p_1 - p_3)^2$.  Also, the spin four-vector for the top quark
may be written as~\cite{st_refs}
\begin{eqnarray}
  s_t^\mu = \left(
    \frac{\vec{p}_t\cdot\hat{s}_t}{m_t},
    \hat{s}_t+\frac{\left(\vec{p}_t\cdot\hat{s}_t\right) \vec{p}_t}{m_t\left(E_t+m_t\right)}
  \right), \nonumber
\end{eqnarray}
where $\vec{p}_t$ and $E_t$ represent the top quark's
three-momentum and energy (in a given reference frame), and where
$\hat{s}_t$ is a unit vector defined in the rest frame of the top
quark.  Note that $p_t \cdot s_t = 0$, as expected.  The symbol
$\overline \Sigma$ denotes that spin and color are averaged for the
inital state and summed for the final state, except for the spin of
the final-state top quark. In each case,
\begin{equation*}
\overline \Sigma \, |{\cal M}|^2 \quad = \quad 
\overline \Sigma \, |{\cal M}_{SM}|^2 
\;\;+\;\;
\overline \Sigma \, |{\cal M}_{NP}|^2 
\;\;+\;\;
2 \mbox{Re}\left(\overline \Sigma {\cal M}_{NP}^{\dagger}\,{\cal M}_{SM}\right) \,.
\end{equation*}

Finally, we adopt the short-hand notation 
$\epsilon(p_1, p_2, p_4, s_t) \equiv 
\epsilon^{\alpha \beta \lambda \eta} \, p_{1\alpha} \, p_{2\beta} \, p_{4\lambda} \, s_{t\eta}$, 
taking $\epsilon^{0123} = +1$.

\setcounter{equation}{0}
%%%%%%%%%%%%%%%%%%%%%%%%%%%%%%%%%%%%%%%%%%%%%%%%%%%%%%%%%%%%%%%%%%%%%%
\section{$s\text{-channel}$ \;:\; 
$\bar b(p_1) \quad c(p_2) \;\; \longrightarrow \;\; t(p_3, s_t) \quad \bar b(p_4)$}
\label{sec:app1}
%%%%%%%%%%%%%%%%%%%%%%%%%%%%%%%%%%%%%%%%%%%%%%%%%%%%%%%%%%%%%%%%%%%%%%
\vspace*{-30pt}
% |M|^2
\begin{flalign*}
\overline \Sigma \, |{\cal M}_{SM}|^2
\quad &= \quad
16\, \GF^2 \, \MW^4 \, \dfrac{|\Vtb|^2 \, |\Vcb|^2}{(Q^2 - \MW^2)^2 + \GW^2 \MW^2} \;
\Big[(p_2 \cdot p_4)\;p_1 \cdot (p_3- m_t s_t)\Big] &&
\end{flalign*}
\begin{flalign*}
\overline \Sigma \, |{\cal M}_{NP}|^2
\quad &= \quad
16 \, \GF^2 \, |\Vcb|^2 \, |\Vtb|^2 \, \nonumber\\
&\mspace{30mu}
\times \bigg[\; 
(p_1 \cdot p_3)(p_2 \cdot p_4) \;(\abbp + \abbm) 
\;+\; 
(p_1 \cdot p_4)(p_2 \cdot p_3) \; (\acp + \acm) \nonumber \\
&\mspace{60mu}
\;+\; 
(p_1 \cdot p_2)(p_3 \cdot p_4) \; (\abp + \abm)
\;-\; 
m_t (p_2 \cdot p_4)(p_1 \cdot s_t) \; (\abbp - \abbm)\nonumber \\
&\mspace{60mu}
\;-\; 
m_t(p_1 \cdot p_4)(p_2 \cdot s_t) \; (\acp - \acm)
\;-\; 
m_t(p_1 \cdot p_2)(p_4 \cdot s_t) \; (\abp - \abm)\nonumber \\
&\mspace{195mu}
\;-\;
8 m_t \mbox{Im}\left(X^T_{LL}X^{S*}_{LL}+X^T_{RR}X^{S*}_{RR}\right)
\epsilon\left(p_1, p_2, p_4, s_t\right)
\;\bigg] &&
\end{flalign*}
\begin{flalign*}
2 \mbox{Re}\left(\overline \Sigma {\cal M}_{NP}^{\dagger}\,{\cal M}_{SM}\right)
\quad &= \quad
64 \, \GF^2 \, \MW^2 \, |\Vtb|^2 |\Vcb|^2
\dfrac{\Big[ \mbox{Re}(\XVLL)(Q^2-\MW^2) \,-\, \mbox{Im}(\XVLL)\GW\MW \Big]}{(Q^2 - \MW^2)^2 + \GW^2 \MW^2} \nonumber \\
&\mspace{200mu} \times \;\Big[(p_2 \cdot p_4)\;p_1 \cdot (p_3 - m_t s_t)\Big] &&
\end{flalign*}

%%%%%%%%%%%%%%%%%%%%%%%%%%%%%%%%%%%%%%%%%%%%%%%%%%%%%%%%%%%%%%%%%%%%%%
\section{$t\text{-channel}$ \;:\;
$b(p_1) \quad c(p_2) \;\; \longrightarrow \;\; t(p_3, s_t) \quad b(p_4)$ \quad \cite{DC-KS}}
\label{sec:app2}
%%%%%%%%%%%%%%%%%%%%%%%%%%%%%%%%%%%%%%%%%%%%%%%%%%%%%%%%%%%%%%%%%%%%%%
\vspace*{-30pt}
% |M|^2
\begin{flalign*}
\overline \Sigma \, |{\cal M}_{SM}|^2
\quad &= \quad
16 \, \GF^2 \, \MW^4 \, \dfrac{|\Vtb|^2\,|\Vcb|^2}{(q^2 - \MW^2)^2} \;
\Big[(p_1 \cdot p_2)\;p_4 \cdot \left(p_3-m_t s_t\right)\Big] &&
\end{flalign*}
\begin{flalign*}
\overline \Sigma \, |{\cal M}_{NP}|^2
\quad &= \quad
16 \, \GF^2 \, |\Vcb|^2 \, |\Vtb|^2 \, \nonumber\\
&\mspace{30mu}
\times \bigg[\; 
(p_1 \cdot p_2)(p_3 \cdot p_4) \;(\abbp + \abbm)
\;+\; 
(p_1 \cdot p_4)(p_2 \cdot p_3) \; (\acp + \acm)\nonumber \\
&\mspace{60mu}
\;+\; 
(p_1 \cdot p_3)(p_2 \cdot p_4) \; (\abp + \abm)
\;-\; 
m_t (p_1 \cdot p_2)(p_4 \cdot s_t) \; (\abbp - \abbm)\nonumber \\
&\mspace{60mu}
\;-\; 
m_t(p_1 \cdot p_4)(p_2 \cdot s_t) \; (\acp - \acm)
\;-\; 
m_t(p_2 \cdot p_4)(p_1 \cdot s_t) \; (\abp - \abm)\nonumber \\
&\mspace{195mu}
\;+\;
8 m_t \mbox{Im}\left(X^T_{LL}X^{S*}_{LL}+X^T_{RR}X^{S*}_{RR}\right)
\epsilon\left(p_1, p_2, p_4, s_t\right)
\;\bigg] &&
\end{flalign*}
\begin{flalign*}
2 \mbox{Re}\left(\overline \Sigma {\cal M}_{NP}^{\dagger}\,{\cal M}_{SM}\right)
%%%%%%%%%%%%%%
\quad &= \quad
64 \, \GF^2 \, \MW^2 \, 
\left(\dfrac{|\Vtb|^2 |\Vcb|^2 \, \mbox{Re}(\XVLL)}{q^2 - \MW^2}\right) \,
\Big[(p_1 \cdot p_2)\;p_4 \cdot \left(p_3 - m_t s_t \right)\Big] &&
\end{flalign*}

%%%%%%%%%%%%%%%%%%%%%%%%%%%%%%%%%%%%%%%%%%%%%%%%%%%%%%%%%%%%%%%%%%%%%%
\section{$t\text{-channel}$ \;:\;
$b(p_1) \quad \bar b(p_2) \;\; \longrightarrow \;\; t(p_3, s_t) \quad \bar c(p_4)$}
\label{sec:app3}
%%%%%%%%%%%%%%%%%%%%%%%%%%%%%%%%%%%%%%%%%%%%%%%%%%%%%%%%%%%%%%%%%%%%%%
\vspace*{-30pt}
% |M|^2
\begin{flalign*}
\overline \Sigma \, |{\cal M}_{SM}|^2
\quad &= \quad
16 \, \GF^2 \, \MW^4 \, \dfrac{|\Vtb|^2\,|\Vcb|^2}{(q^2 - \MW^2)^2} \;
\Big[(p_1 \cdot p_4)\;p_2 \cdot (p_3 - m_t s_t)\Big] &&
\end{flalign*}
\begin{flalign*}
\overline \Sigma \, |{\cal M}_{NP}|^2
\quad &= \quad
16 \, \GF^2 \, |\Vcb|^2 \, |\Vtb|^2 \, \nonumber\\
&\mspace{30mu}
\times \bigg[\; 
(p_1 \cdot p_4)(p_2 \cdot p_3) \; (\abbp + \abbm)
\;+\; 
(p_1 \cdot p_2)(p_3 \cdot p_4) \; (\acp + \acm) \nonumber \\
&\mspace{60mu}
\;+\; 
(p_1 \cdot p_3)(p_2 \cdot p_4) \; (\abp + \abm)
\;-\; 
m_t (p_1 \cdot p_4)(p_2 \cdot s_t) \; (\abbp - \abbm)\nonumber \\
&\mspace{60mu}
\;-\; 
m_t(p_1 \cdot p_2)(p_4 \cdot s_t) \; (\acp - \acm)
\;-\; 
m_t(p_2 \cdot p_4)(p_1 \cdot s_t) \; (\abp - \abm)\nonumber \\
&\mspace{195mu}
\;-\;
8 m_t \mbox{Im}\left(X^T_{LL}X^{S*}_{LL}+X^T_{RR}X^{S*}_{RR}\right)
\epsilon\left(p_1, p_2, p_4, s_t\right)
\;\bigg] &&
\end{flalign*}
\begin{flalign*}
2 \mbox{Re}\left(\overline \Sigma {\cal M}_{NP}^{\dagger}\,{\cal M}_{SM}\right)
%%%%%%%%%%%%%%
\quad &= \quad
64 \, \GF^2 \, \MW^2 \, 
\left(\dfrac{|\Vtb|^2 |\Vcb|^2 \, \mbox{Re}(\XVLL)}{q^2 - \MW^2}\right) \,
\Big[(p_1 \cdot p_4)\;p_2 \cdot (p_3 - m_t s_t)\Big] &&
\end{flalign*}

\end{appendices}

\begin{small}

\end{small}


\begin{thebibliography}{99}

%\cite{Aaltonen:2017efp}
\bibitem{Aaltonen:2017efp} 
  T.~A.~Aaltonen {\it et al.} [CDF and D0 Collaborations],
  %``Combined Forward-Backward Asymmetry Measurements in Top-Antitop Quark Production at the Tevatron,''
  Phys.\ Rev.\ Lett.\  {\bf 120}, 042001 (2018)
  %doi:10.1103/PhysRevLett.120.042001
  [arXiv:1709.04894].
  %%CITATION = doi:10.1103/PhysRevLett.120.042001;%%
  %5 citations counted in INSPIRE as of 10 May 2018
 
\bibitem{RK}
%\cite{Aaij:2014ora}
%\bibitem{Aaij:2014ora} 
  R.~Aaij {\it et al.} [LHCb Collaboration],
  %``Test of lepton universality using $B^{+}\rightarrow K^{+}\ell^{+}\ell^{-}$ decays,''
  Phys.\ Rev.\ Lett.\  {\bf 113}, 151601 (2014),
  %doi:10.1103/PhysRevLett.113.151601
  [arXiv:1406.6482].
  %%CITATION = doi:10.1103/PhysRevLett.113.151601;%%
  %561 citations counted in INSPIRE as of 10 May 2018

\bibitem{P5}
%\cite{Aaij:2015oid}
%\bibitem{Aaij:2015oid} 
  R.~Aaij {\it et al.} [LHCb Collaboration],
  %``Angular analysis of the $B^{0} \to K^{*0} \mu^{+} \mu^{-}$ decay using 3 fb$^{-1}$ of integrated luminosity,''
  JHEP {\bf 1602}, 104 (2016),
  %doi:10.1007/JHEP02(2016)104
  [arXiv:1512.04442];\\
  %%CITATION = doi:10.1007/JHEP02(2016)104;%%
  %300 citations counted in INSPIRE as of 10 May 2018
%
%\cite{Abdesselam:2016llu}
%\bibitem{Abdesselam:2016llu} 
  A.~Abdesselam {\it et al.} [Belle Collaboration],
  %``Angular analysis of $B^0 \to K^\ast(892)^0 \ell^+ \ell^-$,''
  arXiv:1604.04042.
  %%CITATION = ARXIV:1604.04042;%%
  %100 citations counted in INSPIRE as of 10 May 2018

\bibitem{phimumu}  
%\cite{Aaij:2015esa}
%\bibitem{Aaij:2015esa} 
  R.~Aaij {\it et al.} [LHCb Collaboration],
  %``Angular analysis and differential branching fraction of the decay $B^0_s\to\phi\mu^+\mu^-$,''
  JHEP {\bf 1509}, 179 (2015),
  %doi:10.1007/JHEP09(2015)179
  [arXiv:1506.08777].
  %%CITATION = doi:10.1007/JHEP09(2015)179;%%
  %166 citations counted in INSPIRE as of 10 May 2018

\bibitem{RD}  
%\cite{Lees:2013uzd}
%\bibitem{Lees:2013uzd} 
  J.~P.~Lees {\it et al.} [BaBar Collaboration],
  %``Measurement of an Excess of $\bar{B} \to D^{(*)}\tau^- \bar{\nu}_\tau$ Decays and Implications for Charged Higgs Bosons,''
  Phys.\ Rev.\ D {\bf 88}, 072012 (2013),
  %doi:10.1103/PhysRevD.88.072012
  [arXiv:1303.0571];\\
  %%CITATION = doi:10.1103/PhysRevD.88.072012;%%
  %355 citations counted in INSPIRE as of 10 May 2018
%
%\cite{Aaij:2015yra}
%\bibitem{Aaij:2015yra} 
  R.~Aaij {\it et al.} [LHCb Collaboration],
  %``Measurement of the ratio of branching fractions $\mathcal{B}(\bar{B}^0 \to D^{*+}\tau^{-}\bar{\nu}_{\tau})/\mathcal{B}(\bar{B}^0 \to D^{*+}\mu^{-}\bar{\nu}_{\mu})$,''
  Phys.\ Rev.\ Lett.\  {\bf 115}, 111803 (2015)
  Erratum: [Phys.\ Rev.\ Lett.\  {\bf 115}, 159901 (2015)]
  %doi:10.1103/PhysRevLett.115.159901, 10.1103/PhysRevLett.115.111803
  [arXiv:1506.08614].
  %%CITATION = doi:10.1103/PhysRevLett.115.159901, 10.1103/PhysRevLett.115.111803;%%
  %387 citations counted in INSPIRE as of 10 May 2018

% % % % Our earlier papers
\bibitem{tbbc_sofar}
%\cite{Kiers:2011sv}
%\bibitem{Kiers:2011sv} 
  K.~Kiers, T.~Knighton, D.~London, M.~Russell, A.~Szynkman and K.~Webster,
  %``Using t -> b \bar{b} c to Search for New Physics,''
  Phys.\ Rev.\ D {\bf 84}, 074018 (2011),
  %doi:10.1103/PhysRevD.84.074018
  [arXiv:1107.0754];\\
  %%CITATION = doi:10.1103/PhysRevD.84.074018;%%
  %9 citations counted in INSPIRE as of 22 Aug 2017
%\cite{Kiers:2014uqa}
%\bibitem{Kiers:2014uqa} 
  K.~Kiers, P.~Saha, A.~Szynkman, D.~London, S.~Judge and J.~Melendez,
  %``Search for New Physics in Rare Top Decays: $t \bar t$ Spin Correlations and Other Observables,''
  Phys.\ Rev.\ D {\bf 90}, 094015 (2014),
  %doi:10.1103/PhysRevD.90.094015
  [arXiv:1407.1724];\\
  %%CITATION = doi:10.1103/PhysRevD.90.094015;%%
  %6 citations counted in INSPIRE as of 22 Aug 2017
%\cite{Saha:2014vqa}
%\bibitem{Saha:2014vqa} 
  P.~Saha, K.~Kiers, D.~London and A.~Szynkman, 
  %``Detecting New Physics in Rare Top Decays at the LHC,''
  Phys.\ Rev.\ D {\bf 90}, 094016 (2014),
  %doi:10.1103/PhysRevD.90.094016
  [arXiv:1407.1725];\\
  %%CITATION = doi:10.1103/PhysRevD.90.094016;%%
  %4 citations counted in INSPIRE as of 22 Aug 2017
%\cite{Saha:2015lna}
%\bibitem{Saha:2015lna} 
  P.~Saha, K.~Kiers, B.~Bhattacharya, D.~London, A.~Szynkman and J.~Melendez, 
  %``Measuring CP-Violating Observables in Rare Top Decays at the LHC,''
  Phys.\ Rev.\ D {\bf 93}, 054044 (2016),
  %doi:10.1103/PhysRevD.93.054044
  [arXiv:1510.00204].
  %%CITATION = doi:10.1103/PhysRevD.93.054044;%%
  %2 citations counted in INSPIRE as of 22 Aug 2017

\bibitem{stp_old}
%\cite{Han:1998tp}
%\bibitem{Han:1998tp} 
  T.~Han, M.~Hosch, K.~Whisnant, B.~L.~Young and X.~Zhang,
  %``Single top quark production via FCNC couplings at hadron colliders,''
  Phys.\ Rev.\ D {\bf 58}, 073008 (1998),
  %doi:10.1103/PhysRevD.58.073008
  [hep-ph/9806486];\\
  %%CITATION = doi:10.1103/PhysRevD.58.073008;%%
  %110 citations counted in INSPIRE as of 21 Dec 2017
%
%\cite{Tait:2000sh}
%\bibitem{Tait:2000sh} 
  T.~M.~P.~Tait and C.-P.~Yuan,
  %``Single top quark production as a window to physics beyond the standard model,''
  Phys.\ Rev.\ D {\bf 63}, 014018 (2000),
  %doi:10.1103/PhysRevD.63.014018
  [hep-ph/0007298];\\
  %%CITATION = doi:10.1103/PhysRevD.63.014018;%%
  %382 citations counted in INSPIRE as of 21 Dec 2017
%
%\cite{AguilarSaavedra:2008gt}
%\bibitem{AguilarSaavedra:2008gt} 
  J.~A.~Aguilar-Saavedra,
  %``Single top quark production at LHC with anomalous Wtb couplings,''
  Nucl.\ Phys.\ B {\bf 804}, 160 (2008),
  %doi:10.1016/j.nuclphysb.2008.06.013
  [arXiv:0803.3810].
  %%CITATION = doi:10.1016/j.nuclphysb.2008.06.013;%%
  %127 citations counted in INSPIRE as of 21 Dec 2017

\bibitem{stp_new}
%\cite{Goldouzian:2016ufu}
%\bibitem{Goldouzian:2016ufu} 
  R.~Goldouzian and B.~Clerbaux,
  %``Search for anomalous tq$\gamma$ FCNC Couplings in direct single top quark production at the LHC,''
  [arXiv:1612.01795];\\
  %%CITATION = ARXIV:1612.01795;%%
%
%\cite{Pinna:2017tay}
%\bibitem{Pinna:2017tay} 
  D.~Pinna, A.~Zucchetta, M.~R.~Buckley and F.~Canelli,
  %``Single top quarks and dark matter,''
  Phys.\ Rev.\ D {\bf 96}, 035031 (2017),
  %doi:10.1103/PhysRevD.96.035031
  [arXiv:1701.05195];\\
  %%CITATION = doi:10.1103/PhysRevD.96.035031;%%
  %4 citations counted in INSPIRE as of 21 Dec 2017
%
%\cite{Jiang:2017oqi}
%\bibitem{Jiang:2017oqi} 
  X.~Y.~Jiang, J.~Z.~Han, G.~Yang and C.~X.~Yu,
  %``Higgs and Single Top Associated Production at the LHC in the Left-Right Twin Higgs Model,''
  Commun.\ Theor.\ Phys.\  {\bf 67}, 415 (2017);\\
  %doi:10.1088/0253-6102/67/4/415
  %%CITATION = doi:10.1088/0253-6102/67/4/415;%%
%
%\cite{Ahmed:2017cld}
%\bibitem{Ahmed:2017cld} 
  I.~Ahmed,
  %``Charged Higgs Observability in the s-channel Single top Production at LHC,''
  [arXiv:1711.08348];\\
  %%CITATION = ARXIV:1711.08348;%%
%
%\cite{Pani:2017qyd}
%\bibitem{Pani:2017qyd} 
  P.~Pani and G.~Polesello,
  %``Dark matter production in association with a single top quark at the LHC in a two Higgs doublet model with a pseudoscalar mediator,''
  [arXiv:1712.03874].
  %%CITATION = ARXIV:1712.03874;%%
  %1 citations counted in INSPIRE as of 21 Dec 2017
  
\bibitem{four-fermion}
%\bibitem{Ferreira:2005dr}
 P.~M.~Ferreira, O.~Oliveira and R.~Santos,
 %``Flavor changing strong interaction effects on top quark physics at the LHC,''
 Phys.\ Rev.\ D {\bf 73}, 034011 (2006),
 %doi:10.1103/PhysRevD.73.034011
 [hep-ph/0510087];\\
 %%CITATION = doi:10.1103/PhysRevD.73.034011;%%
 %35 citations counted in INSPIRE as of 21 Nov 2017
%\bibitem{Ferreira:2006xe}
 P.~M.~Ferreira and R.~Santos,
 %``Strong flavor changing effective operator contributions to single top quark production,''
 Phys.\ Rev.\ D {\bf 73}, 054025 (2006),
 %doi:10.1103/PhysRevD.73.054025
 [hep-ph/0601078];\\
 %%CITATION = doi:10.1103/PhysRevD.73.054025;%%
 %34 citations counted in INSPIRE as of 21 Nov 2017
%\bibitem{Ferreira:2006in}
 P.~M.~Ferreira and R.~Santos,
 %``Contributions from dimension six strong flavor changing operators to t anti-t, t plus gauge boson, and t plus Higgs boson production at the LHC,''
 Phys.\ Rev.\ D {\bf 74}, 014006 (2006),
 %doi:10.1103/PhysRevD.74.014006
 [hep-ph/0604144];\\
 %%CITATION = doi:10.1103/PhysRevD.74.014006;%%
 %17 citations counted in INSPIRE as of 21 Nov 2017
%\bibitem{Coimbra:2008qp}
 R.~A.~Coimbra, P.~M.~Ferreira, R.~B.~Guedes, O.~Oliveira, A.~Onofre, R.~Santos and M.~Won,
 %``Dimension six FCNC operators and top production at the LHC,''
 Phys.\ Rev.\ D {\bf 79}, 014006 (2009),
 %doi:10.1103/PhysRevD.79.014006
 [arXiv:0811.1743];\\
 %%CITATION = doi:10.1103/PhysRevD.79.014006;%%
 %33 citations counted in INSPIRE as of 21 Nov 2017
%\bibitem{AguilarSaavedra:2010zi}
 J.~A.~Aguilar-Saavedra,
 %``Effective four-fermion operators in top physics: A Roadmap,''
 Nucl.\ Phys.\ B {\bf 843}, 638 (2011),
 Erratum: [Nucl.\ Phys.\ B {\bf 851}, 443 (2011)],
 %doi:10.1016/j.nuclphysb.2011.06.003, 10.1016/j.nuclphysb.2010.10.015
 [arXiv:1008.3562];\\
 %%CITATION = doi:10.1016/j.nuclphysb.2011.06.003, 10.1016/j.nuclphysb.2010.10.015;%%
 %91 citations counted in INSPIRE as of 21 Nov 2017
%\bibitem{Durieux:2014xla}
 G.~Durieux, F.~Maltoni and C.~Zhang,
 %``Global approach to top-quark flavor-changing interactions,''
 Phys.\ Rev.\ D {\bf 91}, 074017 (2015),
 %doi:10.1103/PhysRevD.91.074017
 [arXiv:1412.7166];\\
 %%CITATION = doi:10.1103/PhysRevD.91.074017;%%
 %44 citations counted in INSPIRE as of 21 Nov 2017
%\cite{Aguilar-Saavedra:2017nik}
%\bibitem{Aguilar-Saavedra:2017nik} 
  J.~A.~Aguilar-Saavedra, C.~Degrande and S.~Khatibi,
  %``Single top polarisation as a window to new physics,''
  Phys.\ Lett.\ B {\bf 769}, 498 (2017),
  %doi:10.1016/j.physletb.2017.04.023
  [arXiv:1701.05900].
  %%CITATION = doi:10.1016/j.physletb.2017.04.023;%%
  %2 citations counted in INSPIRE as of 01 Dec 2017

  
  
% % % % ALTAS t-channel 7, 8 TeV  
\bibitem{ATLAS_t7}
%\cite{Aad:2014fwa}
%\bibitem{Aad:2014fwa} 
  G.~Aad {\it et al.} [ATLAS Collaboration],
  %``Comprehensive measurements of $t$-channel single top-quark production cross sections at $\sqrt{s} = 7$
  %TeV with the ATLAS detector,''
  Phys.\ Rev.\ D {\bf 90}, 112006 (2014),
  %doi:10.1103/PhysRevD.90.112006
  [arXiv:1406.7844].
  %%CITATION = doi:10.1103/PhysRevD.90.112006;%%
  %80 citations counted in INSPIRE as of 22 Aug 2017

\bibitem{ATLAS_t8}
% % % % %\cite{ATLAS:2014dja}
% % % % %\bibitem{ATLAS:2014dja} 
% % % %   The ATLAS collaboration,
% % % %   %``Measurement of the Inclusive and Fiducial Cross-Section of Single Top-Quark $t$-Channel Events in $pp$
% % % %   %Collisions at $\sqrt{s}$ = 8 TeV,''
% % % %   ATLAS-CONF-2014-007.
% % % %   %%CITATION = ATLAS-CONF-2014-007;%%
% % % %   %42 citations counted in INSPIRE as of 22 Aug 2017
% % % % 
%\cite{Aaboud:2017pdi}
%\bibitem{Aaboud:2017pdi} 
  M.~Aaboud {\it et al.} [ATLAS Collaboration],
  %``Fiducial, total and differential cross-section measurements of $t$-channel single top-quark production in $pp$ collisions at 8 TeV using data collected by the ATLAS detector,''
  Eur.\ Phys.\ J.\ C {\bf 77}, 531 (2017),
  %doi:10.1140/epjc/s10052-017-5061-9
  [arXiv:1702.02859].
  %%CITATION = doi:10.1140/epjc/s10052-017-5061-9;%%
  %14 citations counted in INSPIRE as of 01 Dec 2017


% % % % ATLAS s-channel 7, 8 TeV  
\bibitem{ATLAS_s7}
%\cite{ATLAS:2011aia}
%\bibitem{ATLAS:2011aia} 
  The ATLAS Collaboration,
  %``Search for s-Channel Single Top-Quark Production in $pp$ Collisions at $\sqrt{s}$ = 7 TeV,''
  ATLAS-CONF-2011-118.
  %%CITATION = ATLAS-CONF-2011-118;%%
  %61 citations counted in INSPIRE as of 22 Aug 2017

\bibitem{ATLAS_s8}
%\cite{Aad:2015upn}
%\bibitem{Aad:2015upn} 
  G.~Aad {\it et al.} [ATLAS Collaboration],
  %``Evidence for single top-quark production in the $s$-channel in proton-proton collisions at $\sqrt{s}=$8 
  %TeV with the ATLAS detector using the Matrix Element Method,''
  Phys.\ Lett.\ B {\bf 756}, 228 (2016),
  %doi:10.1016/j.physletb.2016.03.017
  [arXiv:1511.05980];
  %%CITATION = doi:10.1016/j.physletb.2016.03.017;%%
  %26 citations counted in INSPIRE as of 22 Aug 2017
%\cite{Aad:2014aia}
%\bibitem{Aad:2014aia} 
  %G.~Aad {\it et al.} [ATLAS Collaboration],
  %``Search for $s$-channel single top-quark production in proton–proton collisions at $\sqrt s=8$ TeV with
  %the ATLAS detector,''
  Phys.\ Lett.\ B {\bf 740}, 118 (2015),
  %doi:10.1016/j.physletb.2014.11.042
  [arXiv:1410.0647].
  %%CITATION = doi:10.1016/j.physletb.2014.11.042;%%
  %31 citations counted in INSPIRE as of 22 Aug 2017

% % % % CMS t-channel 7, 8 TeV  
\bibitem{CMS_t7}
%\cite{Chatrchyan:2012ep}
%\bibitem{Chatrchyan:2012ep} 
  S.~Chatrchyan {\it et al.} [CMS Collaboration],
  %``Measurement of the single-top-quark $t$-channel cross section in $pp$ collisions at $\sqrt{s}=7$ TeV,''
  JHEP {\bf 1212}, 035 (2012),
  %doi:10.1007/JHEP12(2012)035
  [arXiv:1209.4533].
  %%CITATION = doi:10.1007/JHEP12(2012)035;%%
  %199 citations counted in INSPIRE as of 22 Aug 2017

\bibitem{CMS_t8}
%\cite{Khachatryan:2014iya}
%\bibitem{Khachatryan:2014iya} 
  V.~Khachatryan {\it et al.} [CMS Collaboration],
  %``Measurement of the t-channel single-top-quark production cross section and of the $\mid V_{tb} \mid$
  %CKM matrix element in pp collisions at $\sqrt{s}$= 8 TeV,''
  JHEP {\bf 1406}, 090 (2014),
  %doi:10.1007/JHEP06(2014)090
  [arXiv:1403.7366].
  %%CITATION = doi:10.1007/JHEP06(2014)090;%%
  %128 citations counted in INSPIRE as of 22 Aug 2017

% % % % CMS s-channel 7, 8 TeV  
\bibitem{CMS_s7n8}
%\cite{Khachatryan:2016ewo}
%\bibitem{Khachatryan:2016ewo} 
  V.~Khachatryan {\it et al.} [CMS Collaboration],
  %``Search for s channel single top quark production in pp collisions at $ \sqrt{s}=7 $ and 8 TeV,''
  JHEP {\bf 1609}, 027 (2016),
  %doi:10.1007/JHEP09(2016)027
  [arXiv:1603.02555].
  %%CITATION = doi:10.1007/JHEP09(2016)027;%%
  %22 citations counted in INSPIRE as of 22 Aug 2017

% % % % ATLAS t-channel 13 TeV  
\bibitem{ATLAS_t13}
%\cite{Aaboud:2016ymp}
%\bibitem{Aaboud:2016ymp} 
  M.~Aaboud {\it et al.} [ATLAS Collaboration],
  %``Measurement of the inclusive cross-sections of single top-quark and top-antiquark $t$-channel production
  %in $pp$ collisions at $\sqrt{s}$ = 13 TeV with the ATLAS detector,''
  JHEP {\bf 1704}, 086 (2017),
  %doi:10.1007/JHEP04(2017)086
  [arXiv:1609.03920].
  %%CITATION = doi:10.1007/JHEP04(2017)086;%%
  %16 citations counted in INSPIRE as of 22 Aug 2017

% % % % CMS t-channel 13 TeV  
\bibitem{CMS_t13}
%\cite{Sirunyan:2016cdg}
%\bibitem{Sirunyan:2016cdg} 
  A.~M.~Sirunyan {\it et al.} [CMS Collaboration],
  %``Cross section measurement of $t$-channel single top quark production in pp collisions at $\sqrt s =$ 13 TeV,''
  Phys.\ Lett.\ B {\bf 772}, 752 (2017),
  %doi:10.1016/j.physletb.2017.07.047
  [arXiv:1610.00678].
  %%CITATION = doi:10.1016/j.physletb.2017.07.047;%%
  %19 citations counted in INSPIRE as of 22 Aug 2017

% %  \bibitem{ATLAS_dists}
% % %\cite{Aaboud:2017pdi}
% % %\bibitem{Aaboud:2017pdi} 
% %   M.~Aaboud {\it et al.} [ATLAS Collaboration],
% %   %``Fiducial, total and differential cross-section measurements of $t$-channel single top-quark production in $pp$ collisions at 8 TeV using data collected by the ATLAS detector,''
% %   Eur.\ Phys.\ J.\ C {\bf 77}, 531 (2017)
% %   %doi:10.1140/epjc/s10052-017-5061-9
% %   [arXiv:1702.02859 [hep-ex]].
% %   %%CITATION = doi:10.1140/epjc/s10052-017-5061-9;%%
% %   %24 citations counted in INSPIRE as of 08 May 2018

\bibitem{CMS_dists_8}
%\cite{CMS:2014ika}
%\bibitem{CMS:2014ika} 
  The CMS Collaboration,
  %CMS Collaboration [CMS Collaboration],
  %``Single top t-channel differential cross section at 8 TeV,''
  CMS-PAS-TOP-14-004.
  %%CITATION = CMS-PAS-TOP-14-004;%%
  %20 citations counted in INSPIRE as of 08 May 2018

\bibitem{CMS_dists_13}
%\cite{CMS:2016xnv}
% \bibitem{CMS:2016xnv}
  The CMS Collaboration,
  %CMS Collaboration [CMS Collaboration],
  %``Measurement of the differential cross section for $t$-channel single-top-quark production at $\sqrt{s}=13~\mathrm{TeV}$,''
  CMS-PAS-TOP-16-004.
  %%CITATION = CMS-PAS-TOP-16-004;%%
  %18 citations counted in INSPIRE as of 08 May 2018
  
\bibitem{LHC_top_WG}
LHC Top Working Group, \\
https://twiki.cern.ch/twiki/bin/view/LHCPhysics/LHCTopWGSummaryPlots

\bibitem{MG5}
%\cite{Alwall:2011uj}
%\bibitem{Alwall:2011uj} 
  J.~Alwall, M.~Herquet, F.~Maltoni, O.~Mattelaer and T.~Stelzer,
  %``MadGraph 5 : Going Beyond,''
  JHEP {\bf 1106}, 128 (2011),
%   doi:10.1007/JHEP06(2011)128
  [arXiv:1106.0522], http://madgraph.hep.uiuc.edu.
  %%CITATION = doi:10.1007/JHEP06(2011)128;%%
  %2551 citations counted in INSPIRE as of 20 Dec 2017
  
\bibitem{FR}
%\cite{Alloul:2013bka}
% \bibitem{Alloul:2013bka} 
  A.~Alloul, N.~D.~Christensen, C.~Degrande, C.~Duhr and B.~Fuks,
  %``FeynRules  2.0 - A complete toolbox for tree-level phenomenology,''
  Comput.\ Phys.\ Commun.\  {\bf 185}, 2250 (2014),
%   doi:10.1016/j.cpc.2014.04.012
  [arXiv:1310.1921], http://feynrules.irmp.ucl.ac.be.
  %%CITATION = doi:10.1016/j.cpc.2014.04.012;%%
  %763 citations counted in INSPIRE as of 20 Dec 2017
  
\bibitem{CTEQ}  
%\cite{Pumplin:2002vw}
% \bibitem{Pumplin:2002vw} 
  J.~Pumplin, D.~R.~Stump, J.~Huston, H.~L.~Lai, P.~M.~Nadolsky and W.~K.~Tung,
  %``New generation of parton distributions with uncertainties from global QCD analysis,''
  JHEP {\bf 0207}, 012 (2002),
%   doi:10.1088/1126-6708/2002/07/012
  [hep-ph/0201195], http://hep.pa.msu.edu/cteq/public/cteq6.html.
  %%CITATION = doi:10.1088/1126-6708/2002/07/012;%%
  %5520 citations counted in INSPIRE as of 20 Dec 2017

\bibitem{MSTW}
%\cite{Martin:2009iq}
% \bibitem{Martin:2009iq} 
  A.~D.~Martin, W.~J.~Stirling, R.~S.~Thorne and G.~Watt,
  %``Parton distributions for the LHC,''
  Eur.\ Phys.\ J.\ C {\bf 63}, 189 (2009),
  %doi:10.1140/epjc/s10052-009-1072-5
  [arXiv:0901.0002], https://mstwpdf.hepforge.org.
  %%CITATION = doi:10.1140/epjc/s10052-009-1072-5;%%
  %4024 citations counted in INSPIRE as of 20 Dec 2017

\bibitem{NLO_sch_813}
%\cite{Kidonakis:2010tc}
%\bibitem{Kidonakis:2010tc} 
  N.~Kidonakis,
  %``NNLL resummation for s-channel single top quark production,''
  Phys.\ Rev.\ D {\bf 81}, 054028 (2010),
  %doi:10.1103/PhysRevD.81.054028
  [arXiv:1001.5034].
  %%CITATION = doi:10.1103/PhysRevD.81.054028;%%
  %573 citations counted in INSPIRE as of 23 Aug 2017

\bibitem{NLO_tch_8}  
%\cite{Kidonakis:2012db}
%\bibitem{Kidonakis:2012db} 
  N.~Kidonakis,
  %``Differential and total cross sections for top pair and single-top production,''
  %doi:10.3204/DESY-PROC-2012-02/251
  arXiv:1205.3453;
  %%CITATION = doi:10.3204/DESY-PROC-2012-02/251;%%
  %152 citations counted in INSPIRE as of 23 Aug 2017
%\cite{Kidonakis:2013zqa}
%\bibitem{Kidonakis:2013zqa} 
  %N.~Kidonakis,
  %``Top Quark Production,''
  %doi:10.3204/DESY-PROC-2013-03/Kidonakis
  arXiv:1311.0283.
  %%CITATION = doi:10.3204/DESY-PROC-2013-03/Kidonakis;%%
  %61 citations counted in INSPIRE as of 23 Aug 2017
  
\bibitem{toppol}  
 %\cite{Chatrchyan:2013wua}
 %\bibitem{Chatrchyan:2013wua} 
  S.~Chatrchyan {\it et al.} [CMS Collaboration],
  %``Measurements of $t\bar{t}$ spin correlations and top-quark polarization using dilepton final states in $pp$ collisions at $\sqrt{s}$ = 7 TeV,''
  Phys.\ Rev.\ Lett.\  {\bf 112}, 182001 (2014),
  %doi:10.1103/PhysRevLett.112.182001
  [arXiv:1311.3924];\\
  %%CITATION = doi:10.1103/PhysRevLett.112.182001;%%
  %70 citations counted in INSPIRE as of 24 Aug 2017
 %\cite{Khachatryan:2016xws}
 %\bibitem{Khachatryan:2016xws} 
  V.~Khachatryan {\it et al.} [CMS Collaboration],
  %``Measurements of t t-bar spin correlations and top quark polarization using dilepton final states in pp collisions at sqrt(s) = 8 TeV,''
  Phys.\ Rev.\ D {\bf 93}, 052007 (2016),
  %doi:10.1103/PhysRevD.93.052007
  [arXiv:1601.01107];\\
  %%CITATION = doi:10.1103/PhysRevD.93.052007;%%
  %24 citations counted in INSPIRE as of 24 Aug 2017
  %\cite{Aad:2013ksa}
  %\bibitem{Aad:2013ksa} 
  G.~Aad {\it et al.} [ATLAS Collaboration],
  %``Measurement of Top Quark Polarization in Top-Antitop Events from Proton-Proton Collisions at $\sqrt{s}$ = 7  TeV Using the ATLAS Detector,''
  Phys.\ Rev.\ Lett.\  {\bf 111}, 232002 (2013),
  %doi:10.1103/PhysRevLett.111.232002
  [arXiv:1307.6511].
  %%CITATION = doi:10.1103/PhysRevLett.111.232002;%%
  %59 citations counted in INSPIRE as of 24 Aug 2017

\bibitem{toppol_single}  
 %\cite{Khachatryan:2015dzz}
 %\bibitem{Khachatryan:2015dzz} 
  V.~Khachatryan {\it et al.} [CMS Collaboration],
  %``Measurement of top quark polarisation in t-channel single top quark production,''
  JHEP {\bf 1604}, 073 (2016),
  %doi:10.1007/JHEP04(2016)073
  [arXiv:1511.02138].
  %%CITATION = doi:10.1007/JHEP04(2016)073;%%
  %27 citations counted in INSPIRE as of 24 Aug 2017
 
\bibitem{bcharge} 
%\cite{Abazov:2006vd}
%\bibitem{Abazov:2006vd} 
  V.~M.~Abazov {\it et al.} [D0 Collaboration],
  %``Experimental discrimination between charge 2e/3 top quark and charge 4e/3 exotic quark production scenarios,''
  Phys.\ Rev.\ Lett.\  {\bf 98}, 041801 (2007),
  %doi:10.1103/PhysRevLett.98.041801
  [hep-ex/0608044];
  %%CITATION = doi:10.1103/PhysRevLett.98.041801;%%
  %97 citations counted in INSPIRE as of 10 Oct 2017
%\cite{Aad:2013uza}
%\bibitem{Aad:2013uza} 
  G.~Aad {\it et al.} [ATLAS Collaboration],
  %``Measurement of the top quark charge in $pp$ collisions at $\sqrt{s} =$ 7 TeV with the ATLAS detector,''
  JHEP {\bf 1311}, 031 (2013),
  %doi:10.1007/JHEP11(2013)031
  [arXiv:1307.4568].
  %%CITATION = doi:10.1007/JHEP11(2013)031;%%
  %44 citations counted in INSPIRE as of 10 Oct 2017
  
\bibitem{st_refs}
C.~Itzykson and J.~B.~Zuber, \textit{Quantum Field Theory}, McGraw-Hill (1980),\\ \mbox{ISBN 0-07-032071-3};\\
%
%\cite{Kiers:2006aq}
%\bibitem{Kiers:2006aq} 
  K.~Kiers, A.~Szynkman and D.~London,
  %``CP violation in supersymmetric theories: anti-t(2) ---> anti-t(1) tau- tau+,''
  Phys.\ Rev.\ D {\bf 74}, 035004 (2006),
  %doi:10.1103/PhysRevD.74.035004
  [hep-ph/0605123].
  %%CITATION = doi:10.1103/PhysRevD.74.035004;%%
  %16 citations counted in INSPIRE as of 18 Dec 2017
  
\bibitem{DC-KS}
David Cusick and Kristian Stephens, \textit{private communication}.
  
\end{thebibliography}
\end{document}